\documentclass[manuscript,screen]{acmart}
\usepackage{graphics}
\usepackage{hyperref}
\usepackage{multirow}
\usepackage{soul}
\usepackage{tabularx}
\usepackage{tcolorbox}
\usepackage[normalem]{ulem}
\usepackage{xcolor}
\usepackage{array}
\usepackage{tabularx}
\usepackage{booktabs}
\usepackage{graphicx}

\usepackage{paralist}
\usepackage{enumitem}

\usepackage{bchart}
\usepackage{titlesec}

\usepackage[outercaption]{sidecap}
\usepackage[figuresright]{rotating}

\usepackage{courier}
\usepackage{hyperref}

\usepackage{color}
\usepackage{listings}
\usepackage{xcolor}

\definecolor{codegreen}{rgb}{0,0.6,0}
\definecolor{codegray}{rgb}{0.5,0.5,0.5}
\definecolor{codepurple}{rgb}{0.58,0,0.82}
\definecolor{backcolour}{rgb}{.99,.99,.99}
\definecolor{lightgray}{gray}{0.9}

\definecolor{dkgreen}{rgb}{0,0.6,0}
\definecolor{gray}{rgb}{0.5,0.5,0.5}
\definecolor{mauve}{rgb}{0.58,0,0.82}

\usepackage{xpatch}

\xpatchcmd{\refstepcounter}{%
  \stepcounter{#1}%
}{%
  \stepcounter{#1}%
  \xdef\scr{\number\value{#1}}%
}{\typeout{success}}{\typeout{failure}}

\usepackage{tikz}
\usepackage{todonotes}

\usepackage{listings}
\usepackage{xcolor}
\usepackage{outlines}

\definecolor{codeblue}{rgb}{0.3,0.3,0.9}
\definecolor{codegreen}{rgb}{0,0.6,0}
\definecolor{codegray}{rgb}{0.5,0.5,0.5}
\definecolor{codepurple}{rgb}{0.58,0,0.82}

\lstdefinestyle{javastyle}{
    language=Java,
    basicstyle=\ttfamily\scriptsize,
    keywordstyle=\color{codeblue},
    stringstyle=\color{codegreen},
    commentstyle=\color{codegray},
    morecomment=[l][\color{codepurple}]{\#},
    numbers=left,
    xleftmargin=1em,
    numberstyle=\tiny\color{codegray},
    stepnumber=1,
    numbersep=10pt,
    tabsize=1,
    showspaces=false,
    showstringspaces=false,
    breaklines=true,
    breakatwhitespace=true,
    escapeinside={(*@}{@*)}
}

\usepackage[T1]{fontenc}
\usepackage{lmodern}
\usepackage{textcomp}

\definecolor{1c1}{RGB}{188,162,6}
\definecolor{1c2}{RGB}{137,129,80}
\definecolor{1c3}{RGB}{239,167,31}
\definecolor{1c4}{RGB}{88,194,241}
\definecolor{1c5}{RGB}{6,180,188}

\AtBeginDocument{%
  \providecommand\BibTeX{{%
    \normalfont B\kern-0.5em{\scshape i\kern-0.25em b}\kern-0.8em\TeX}}}

\graphicspath{{}{images/}{dia/}}
\DeclareGraphicsExtensions{.pdf,.png}

\newcounter{o}
\definecolor{1c1}{RGB}{188,162,6}
\definecolor{1c2}{RGB}{137,129,80}
\definecolor{1c3}{RGB}{239,167,31}
\definecolor{1c4}{RGB}{88,194,241}
\definecolor{1c5}{RGB}{6,180,188}

\def\bf{\textbf}

\newcommand{\urls}[1]{{\scriptsize\url{#1}}}

\usepackage{soul}
\usepackage{color}
\usepackage{xcolor}

\newcommand{\pa}[1]{\noindent\textbf{#1}}

\newcommand{\RQOne}{What is the comparative performance of our approach against existing baselines?}
\newcommand{\RQTwo}{How does translation impact the quality and security of the generated code?}

\newcommand{\RQThreeNew}{What is the contribution of each agent to the overall 
effectiveness of \tool{}?}

\newcommand{\NLSpec}{NL-Specification}
\newcommand{\NLSpecS}{NL-Specification\space}

\newcommand{\tool}{BabelCoder}
\newcommand{\toolS}{BabelCoder\space}

\newcommand{\CAGen}{\ensuremath{\text{CA}_{\text{gen}}}}
\newcommand{\CAEval}{\ensuremath{\text{CA}_{\text{eval}}}}

\definecolor{mainColor}{HTML}{000000}    \definecolor{subColor}{HTML}{004400} 
\newtcolorbox{boxFindings}{
    fontupper = \bf\color{mainColor}, % font color
    boxrule = 1.5pt,
    colframe = subColor,
    rounded corners,
    arc = 5pt   % corners roundness
}

\newtcolorbox[auto counter, number within=section]{promptBox}[2][]{colframe=blue!50!black, colback=blue!10, coltitle=black, 
    fonttitle=\bfseries, title=Text Box~\thetcbcounter: #2,#1}

\definecolor{keyword}{rgb}{0.58, 0.0, 0.83}
\definecolor{string}{rgb}{0.2, 0.6, 0.2}
\definecolor{comment}{rgb}{0.5, 0.5, 0.5}
\definecolor{extra}{rgb}{0.58, 0.0, 0.83}
\definecolor{correctMark}{rgb}{0.80, 0.99, 0.80}
\definecolor{wrongMark}{rgb}{0.99, 0.80, 0.80}
\definecolor{targetMark}{rgb}{0.80, 0.80, 0.80}
\definecolor{highlight}{rgb}{0.80, 0.90, 0.80}
\usepackage{array}
\newcolumntype{L}[1]{>{\raggedright\arraybackslash}p{#1}}

\usepackage{cite}
\usepackage{amsmath,amsfonts}
\usepackage{algorithmic}
\usepackage{graphicx}
\usepackage{textcomp}
\usepackage{xcolor}
\usepackage{multirow}

\usepackage[table]{xcolor}
\usepackage{tcolorbox}
\usepackage{xcolor}
\usepackage{listings}

\tcbset{
  mypromptbox/.style={
    colback=gray!5,     % background
    colframe=black!80, % border
    fonttitle=\bfseries,
    fontupper=\footnotesize,
    boxrule=0.5pt,
    arc=4pt,
    left=0pt,
    right=0pt,
    top=0pt,
    bottom=0pt,
  }
}

\begin{document}

\title{\tool: Agentic Code Translation with Specification Alignment}

\author{Fazle Rabbi}
\authornote{Both authors contributed equally to this research.}
\email{fazle.rabbi@mail.concordia.ca}
\orcid{0009-0007-8992-9682}
\affiliation{%
  \institution{Concordia University}
  \city{Montreal}
  \state{QC}
  \country{Canada}
}
\author{Soumit Kanti Saha}
\authornotemark[1]
\email{s_soumit@live.concordia.ca}
\orcid{0009-0000-8771-7829}
\affiliation{%
  \institution{Concordia University}
  \city{Montreal}
  \state{QC}
  \country{Canada}
}
\author{Tri Minh Triet Pham}
\email{p_triet@encs.concordia.ca}
\orcid{0009-0000-4455-2315}
\affiliation{%
  \institution{Concordia University}
  \city{Montreal}
  \state{QC}
  \country{Canada}
}
\author{Song Wang}
\email{wangsong@yorku.ca}
\orcid{0000-0003-0617-2877}
\affiliation{%
  \institution{York University}
  \city{Toronto}
  \state{ON}
  \country{Canada}
}
\author{Jinqiu Yang}
\email{jinqiu.yang@concordia.ca}
\orcid{0000-0003-4282-406X}
\affiliation{%
  \institution{Concordia University}
  \city{Montreal}
  \state{QC}
  \country{Canada}
}
\renewcommand{\shortauthors}{Fazle et al.}
\acmArticleType{Review}
\acmCodeLink{https://github.com/profile/repo}
\acmDataLink{htps://zenodo.org/link}
\acmContributions{.}

\begin{abstract}
As software systems evolve, developers increasingly work across multiple programming languages and often face the need to migrate code from one language to another. While automatic code translation offers a promising solution, it has long remained a challenging task. Recent advancements in Large Language Models (LLMs) have shown potential for this task, yet existing approaches remain limited in accuracy and fail to effectively leverage contextual and structural cues within the code. Prior work has explored translation and repair mechanisms, but lacks a structured, agentic framework where multiple specialized agents collaboratively improve translation quality. In this work, we introduce \tool{}, an agentic framework that performs code translation by decomposing the task into specialized agents for translation, testing, and refinement, each responsible for a specific aspect such as generating code, validating correctness, or repairing errors. We evaluate \tool{} on four benchmark datasets and compare it against four state-of-the-art baselines. \tool{} outperforms existing methods on 26 of the 29 language pairs evaluated, by up to 24.67 percentage points, achieving an average accuracy of 94.18\%.
\end{abstract}



\maketitle

\section{Introduction}
\label{introduction}
Software development now involves a diverse array of programming languages, each tailored to specific application requirements. As technologies and project goals evolve, systems often must migrate to more modern or suitable languages. For example, developers are migrating unsafe C code to Rust for better memory safety~\citep{zhang2023ownership, ling2022rust, emre2021translating}, and porting legacy COBOL code to modern languages to improve maintainability and performance~\citep{sneed2010migrating, gandhi2024translation}. However, manually rewriting large codebases remains time-consuming and error-prone and demands significant developer effort and domain expertise.
% For example, unsafe C code is increasingly being translated into the safer Rust language to enhance memory safety and reliability~\citep{zhang2023ownership, ling2022rust, emre2021translating}. Similarly, legacy code written in low-resource languages such as COBOL is being migrated to modern languages to improve maintainability and performance~\citep{sneed2010migrating, gandhi2024translation}.
% However, manually rewriting large codebases in a new language is a time-consuming and error-prone task that demands significant developer effort and domain expertise.

Recent studies have explored Large Language Models (LLMs) for code translation from multiple perspectives. Early works such as Pan et al.~\citep{pan2024lost} and Yin et al.~\citep{yin2024rectifiercodetranslationcorrector} reported low correctness and inconsistent bugs, showing limited gains from prompt engineering. %To improve translation reliability, Yin et al. [33] proposed Rectifier, a fine-tuned CodeT5+ model for error correction. 
Subsequent systems including UniTrans ~\citep{yang2024exploring}, ExeCoder~\citep{he2025execoder}, and CoTR~\citep{yang2025assessing} enhance robustness by incorporating testing mechanisms. More recent efforts, such as TransAgent ~\citep{yuan2024transagentllmbasedmultiagentcode}, adopt a multi-agent decomposition strategy to improve modularity, whereas SpecTra~\citep{nitin2024spectraenhancingcodetranslation} utilizes multi-modal specifications (e.g., static assertions and I/O examples) to guide translation. InterTrans~\citep{macedo2024intertrans} advances further by performing translation through multiple intermediate representations, achieving state-of-the-art results on benchmarks such as CodeNet, HumanEval-X, and TransCoder.

Despite these advancements, existing approaches share several limitations:
\begin{itemize}
\item \textbf{Limited exploitation of rich representations.} While some approaches use multi-modal cues, {few generate and validate intermediate natural language specifications to guide semantic translation and drive iterative repair. Natural language guidance itself has been explored, for example SpecTra~\citep{nitin2024spectraenhancingcodetranslation} generates multi-modal specifications for a single-pass translation, but such methods do not validate or refine the specification against runtime feedback.}

\item {\textbf{Monolithic or shallow modeling of translation.} Most methods rely on single-step, end-to-end LLM translation, which limits adaptability and interpretability when dealing with diverse or complex code structures~\citep{pan2024lost}.}

\item \textbf{Restricted refinement and validation mechanisms.} Refinement often focuses narrowly on syntactic or variable-level consistency, without leveraging deeper semantic reasoning for correctness validation {\citep{yang2024exploring, yuan2024transagentllmbasedmultiagentcode}}.

\item \textbf{High computational cost and poor scalability.} Multi-path frameworks (e.g., InterTrans) achieve higher accuracy but at the expense of extensive LLM invocations, making them difficult to scale in practice {\citep{macedo2024intertrans}}.
\end{itemize}

These limitations underscore the need for more structured, agent-assisted frameworks that can iteratively refine translations and adapt to diverse contexts. Figure~\ref{fig:babelcoder_motivation_example} shows a challenging example from the EvalPlus benchmark~\citep{liu2024your}, which state-of-the-art approaches fail to translate correctly due to misinterpretation of the underlying logic. In contrast, {\tool} resolves this example correctly using its dedicated Refinement Agent, including iterative fixes.

{As the figure shows, the baseline fails not for lack of Java knowledge but because it loses the source program's intent: without an intermediate specification, it mistranslates both the assertion and the return type. \tool{} addresses this by first generating and \emph{validating} an \NLSpecS{}, and the example concretely illustrates the four limitations above, the \NLSpecS{} supplies the rich, language-agnostic representation missing from single-modal translation (limitation~1), separating translation, validation, and repair across agents replaces monolithic end-to-end generation (limitation~2), validating the \NLSpecS{} against test feedback yields semantic rather than purely syntactic repair (limitation~3), and the bounded, test-terminated refinement loop keeps LLM usage low relative to multi-path search (limitation~4).}

\begin{figure}[]
\centering
\begin{tcolorbox}[mypromptbox, width=0.9\linewidth]
{\footnotesize \underline{Source Code $(HumanEval\_35,\ Python)$}}
\begin{lstlisting}[escapechar=~]
1. def max_element(l: list):
2.     assert all(
3.             type(x) in [int, float] 
4.             for x in l
5.         ), "invalid inputs"
6.     return~\colorbox{targetMark}{max(l)}~
\end{lstlisting}
\end{tcolorbox}

\begin{tcolorbox}[mypromptbox, width=0.9\linewidth]
{\footnotesize Translation $(to\ Java)$ without Refinement Agent \\
\underline{Approach as baselines \citep{pan2024lost}, \citep{yang2024exploring}, \citep{yuan2024transagentllmbasedmultiagentcode}}
}
\begin{lstlisting}[escapechar=~]
...
1. public class HumanEval_35 {
2.     public static Number maxElement
3.             (List<? extends Number> l) {
4.         if (!l.stream().allMatch(x -> 
5.                 x instanceof Integer || 
6.                 x instanceof Double || 
7.                 x instanceof Float)) {
8.             throw new IllegalArgumentException(
9.                     "invalid inputs");
10.        }
11.        return l.stream().
12.               ~\colorbox{wrongMark}{mapToDouble(Number::doubleValue).}~
13.                max().orElseThrow();
14.    }
15. }
\end{lstlisting}
\end{tcolorbox}

\begin{tcolorbox}[mypromptbox, width=0.9\linewidth]
{\footnotesize \underline{\textcolor{blue}{Validated NL-Specification (produced by the Refinement Agent)}}}
\begin{lstlisting}[escapechar=~]
1. FUNCTION max_element(l):
2.     ASSERT every element x in l is an
3.            integer or a float, otherwise
4.            raise ~\colorbox{correctMark}{"invalid inputs"}~
5.     RETURN the ~\colorbox{correctMark}{maximum element}~ of l
\end{lstlisting}
\end{tcolorbox}

\begin{tcolorbox}[mypromptbox, width=0.9\linewidth]
{\footnotesize \underline{Translation $(to\ Java)$ with Refinement Agent (Our Approach)}}
\begin{lstlisting}[escapechar=~]
...
1. public class HumanEval_35 {
2.     public static Number maxElement
3.             (List<Number> l) {
4.         for (Number x : l) {
5.             if (!(x instanceof Integer ||
6.                     x instanceof Float ||
7.                     x instanceof Double)) {
8.                 throw new AssertionError(
9.                         "invalid inputs"
10.                );
11.            }
12.        }
13.        Number max = l.get(0);
14.        for (Number x : l) {
15.           ~\colorbox{correctMark}{if(x.doubleValue()>max.doubleValue())}~{
17.                max = x;
18.            }
19.        }
20.        return max;
21.    }
22.}
\end{lstlisting}
\end{tcolorbox}

\caption{A motivating example from the EvalPlus benchmark~\citep{liu2024your}. Prior methods~\citep{yang2024exploring, pan2024lost, yuan2024transagentllmbasedmultiagentcode} translate the Python \texttt{max\_element} directly and misread its logic, mapping the input assertion to an \texttt{IllegalArgumentException} and reducing each element to a \texttt{double} via \texttt{mapToDouble}, so the returned value is no longer the original element. \tool's Refinement Agent instead first derives a \emph{validated} \NLSpecS{} capturing the language-agnostic intent (assert all elements are numeric, then return the maximum element), which guides a correct, contract-preserving translation (Section~\ref{refinement-agent}).}

% \caption{An example from the EvalPlus benchmark~\citep{liu2024your},  where prior methods~\citep{yang2024exploring, pan2024lost} fail to correctly translate a Python max\_element function  to Java due to misinterpreting its logic. \tool's Refinement Agent (Section~\ref{refinement-agent}) generates validated {\NLSpec} that guides accurate translation and bug fixing (Section~\ref{approach}), resulting in a correct and robust output. \color{red} enhance based on Reiew R1.2}
\label{fig:babelcoder_motivation_example}
\Description{babelcoder motivation example}
\end{figure}
In this work, we propose an agentic framework for automated code translation, namely \tool, which first translates the code and then refines it using specifications and tests inferred from the source. \toolS is composed of three agents: a \textit{Translation Agent}, a \textit{Test Agent}, and a \textit{Refinement Agent}. 
The \textit{Translation Agent} generates an initial translation skeleton using source code and further refines the generated code based on the validated \NLSpecS produced by the \textit{Refinement Agent}. The \textit{Test Agent} generates inputs for the source code, derives the corresponding oracles, and evaluates the correctness of the translated code by executing the generated tests when needed.
The \textit{Refinement Agent} autonomously generates enriched specifications to improve semantic understanding during translation. It then applies a multi-step iterative repair process to improve the translated code’s quality.
To address translation-related bugs, our system integrates multiple complementary repair mechanisms that target compilation errors, runtime failures, and assertion violations. These include general-purpose prompt-based correction and Spectrum-Based Fault Localization (SBFL). In addition, we introduce two novel repair strategies. First, \textbf{\NLSpecS Validation} to validate the translated code's behavior. Since \NLSpecS summarizes the source logic in a language-agnostic form, mismatches between it and the output indicate semantic errors, guiding repairs beyond syntactic fixes. Second, \textbf{finding and fixing scopes of bugs using LLM} to focus correction efforts more effectively.
To strengthen the model’s semantic understanding and improve repair accuracy, we incorporate summarized error messages and contextual cues directly into the LLM prompts. This enhanced semantic context helps the model reason more effectively about the intent and behavior of the original code. Through this modular and layered architecture, our framework significantly improves code executability and robustness, while also elevating the quality of code translation beyond existing LLM-based systems.

Our evaluation consists of code translation tasks across 2 LLMs, 4 benchmark datasets, and five programming languages. Although repository-level translation is an important goal for real-world software migration, this work follows prior studies~\citep{pan2024lost, macedo2024intertrans, yuan2024transagentllmbasedmultiagentcode} that focus on function-level~\citep{liu2024your} and file-level~\citep{ahmad-etal-2021-avatar, puri2021codenet} translation tasks. Our scope aligns with the widely-used benchmarks~\citep{ahmad-etal-2021-avatar, puri2021codenet, liu2024your}, which provide paired code snippets instead of full repositories, enabling controlled and comparable evaluation of translation correctness. Translating entire repositories introduces additional challenges, such as preserving architecture, managing cross-file dependencies, and exceeding the context window limits of current LLMs~\citep{wang2024repotransbench}, which often make end-to-end translation infeasible. Nonetheless, our multi-agent architecture is modular and inherently scalable: its Translation, Test, and Refinement Agents can be applied incrementally across files or modules, allowing straightforward extension to repository-level translation in future work.

% \todo{the paragraph on repo-level should be moved here.} \fazle{moved}

To evaluate the effectiveness of our proposed framework, we conduct extensive experiments on four publicly available datasets: AVATAR~\citep{ahmad-etal-2021-avatar}, CodeNet~\citep{puri2021codenet}, EvalPlus~\citep{liu2024your}, and  TransCoder \citep{lachaux2020unsupervised} dataset cleaned by UniTrans~\citep{yang2024exploring}. These datasets cover a broad range of programming problems and five programming languages (Java, C++, C, Go, and Python), allowing us to assess both the generalizability and reliability of our system. We adopt computational accuracy, measured using test case pass rates, as our primary evaluation metric to reflect the functional correctness of translated code. Our framework leverages GPT-4o as the underlying LLM, chosen for its strong performance in code understanding and generation. For baseline comparison, we evaluate against four state-of-the-art approaches: Lost in Translation~\citep{pan2024lost}, UniTrans~\citep{yang2024exploring} and TransAgent~\citep{yuan2024transagentllmbasedmultiagentcode} and InterTrans~\citep{macedo2024intertrans}. Experimental results show that our multi-agent framework consistently outperforms existing methods across all four datasets and five languages, demonstrating higher correctness and robustness in code translation and repair.

Our contribution to this work is as follows:
\begin{enumerate}
    \item We propose a novel multi-agent framework for automatic code translation and bug fixing. Through a comprehensive evaluation against existing approaches, we demonstrate that our method significantly outperforms them in both translation accuracy and reliability.
    
    % \item We utilize both the source code and intermediate natural language specifications (\NLSpec) to improve the quality and correctness of translations across multiple target languages.

    \item We utilize both the source code and intermediate natural language specifications (\NLSpec) to improve the quality and correctness of translations across multiple target languages. {Unlike prior work that uses such specifications only to prompt a single translation, we validate and align the \NLSpecS against the source and treat any mismatch as a semantic repair signal.}
    
    % \item We introduce a \textit{Bug Localization} system to identify the relevant bug region and a corresponding LLM-based fixing that performs targeted repairs based on this contextual information, both integrated within our Refinement Agent.

    \item We introduce a \textit{Bug Localization} system that {combines spectrum-based fault localization with an LLM-based scope estimator} to identify the relevant bug region, and a corresponding LLM-based fixing that performs targeted repairs based on this contextual information, both integrated within our Refinement Agent.
    
    \item { We convert the EvalPlus test cases to Java using our own automated unit test converter} to enable correctness evaluation of the translated code in that language, and we publicly release this dataset to support future research.

    \item { We show through an ablation study, the impact of each agent in BabelCoder, which will help future researchers build on and extend similar studies.}
\end{enumerate}

% To better understand how well our framework works and what factors influence its performance, we came up with the following research questions (RQs). These questions help us break down the problem, explore the strengths and weaknesses of our approach, and guide our experiments.\\
To assess our framework’s effectiveness and influencing factors, we define the following Research Questions (RQs) to guide analysis and experimentation.\\
\begin{enumerate}
    \item \textbf{RQ1.} \RQOne
    \item \textbf{RQ2.} \RQTwo
    \item \textbf{RQ3.} {\RQThreeNew}
\end{enumerate}

All code, datasets, and implementation details of our work are available for reproducibility at\footnote{https://github.com/anonprox/babelcoder}.

\begin{figure*}[]
    \centering
    \includegraphics[width=0.9\textwidth]{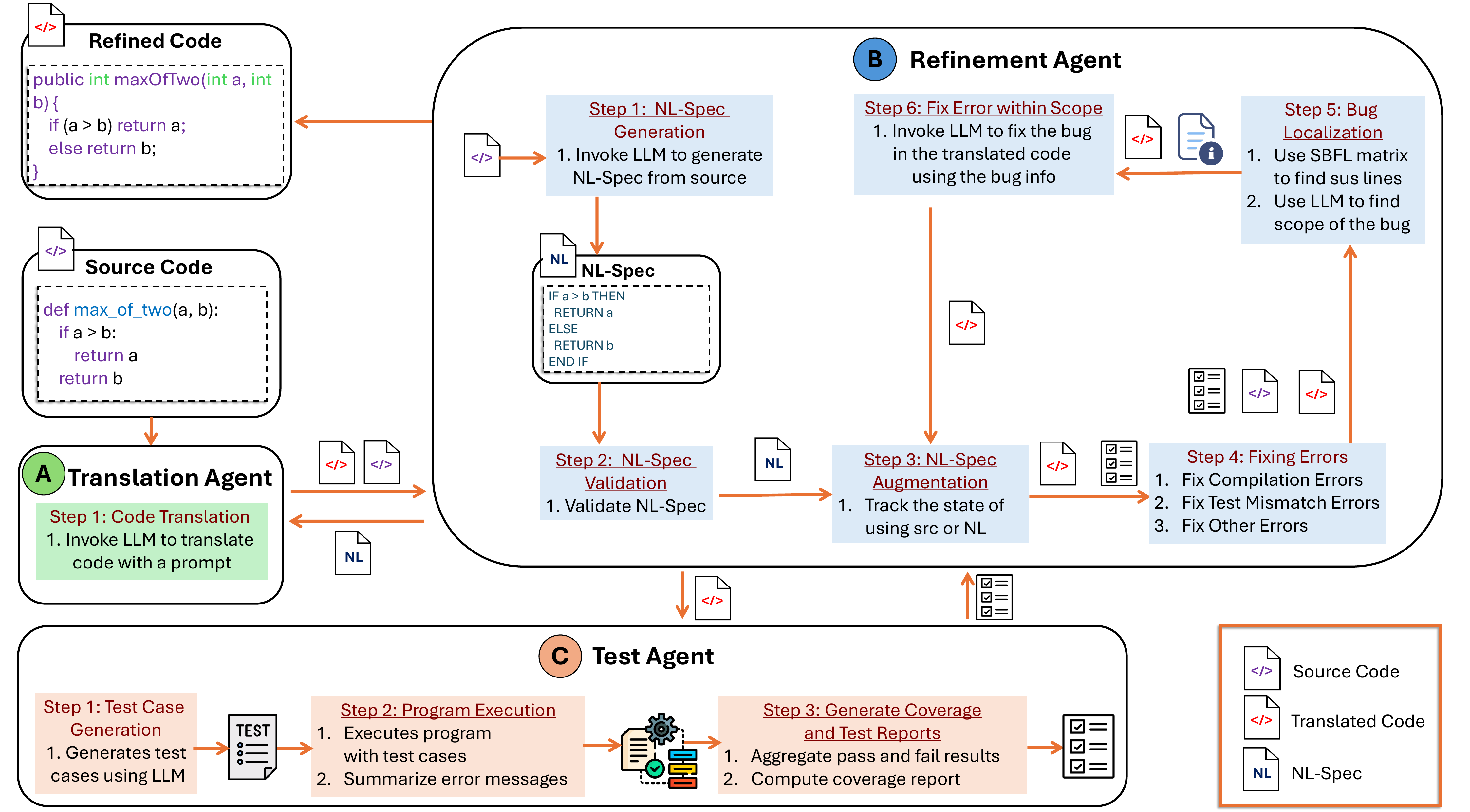}
    \caption{Overview of \toolS}
    \label{fig:overview}
    \Description{overview}
\end{figure*}

\section{The Design of \tool}
\label{approach}

%\noindent\textbf{Overview.}
%\label{system-overview}

Figure~\ref{fig:overview} presents the overall architecture of \tool, including three major components: the \textit{A. Translation Agent}, the \textit{B. Refinement Agent}, and the \textit{C. Test Agent}.

{A translation begins when \toolS{} passes the source code to the Translation Agent (Agent A). Agent A produces an initial translation from the source code and forwards it to the Refinement Agent (Agent B). Agent B then drives a test-driven refinement loop, and its interactions with the other agents are determined by that loop. In every iteration, Agent B first calls the Test Agent (Agent C) to compile the current translation, run the generated tests, and summarize the resulting errors together with coverage. If all tests pass, Agent B stops and returns the current code. If some tests fail, Agent B takes one of two actions before the next iteration, and the state machine determines which one. When the current input configuration has stopped increasing the number of passing tests, Agent B requests a new translation from Agent A, and the state machine advances the input for the next attempt to the source code, the \NLSpecS generated and validated from the source (Section~\ref{nl-validation}), or both (Section~\ref{nl-agumentation}). Otherwise, Agent B repairs the current translation in place, localizing the fault with Spectrum-Based Fault Localization and an LLM-based scope estimator (Section~\ref{bug-scope-estimation}) and applying a targeted fix. The loop repeats until all generated tests pass or a preset maximum number of refinement cycles is reached, and Agent B then outputs the final translation.}

{We use the term agent in the operational sense adopted in recent work on LLM agents~\citep{yao2023react, shinn2023reflexion, wang2024survey}. An agent is an LLM-driven component that uses external tools, observes feedback from its environment, and makes autonomous and state-dependent decisions rather than following a fixed script. Under this definition BabelCoder is not a static workflow, for two reasons. First, its control flow is decided at run time from test feedback. The state machine selects the translation input form, the localization component selects a repair scope, and the loop decides whether to re-translate, which repair to apply, and when to stop, and each of these choices depends on the observed test outcomes. As a result, different inputs follow different paths through the same set of agents. Second, the agents themselves make open-ended decisions that a fixed rule engine cannot. The bug-scope agent classifies a fault into one of several code scopes, and the repair agent formulates and applies the fix, and both are driven by the LLM. The agents also use tools such as the compiler, the test runner, and the coverage and fault-localization machinery, and they exchange intermediate artifacts. These properties are what we mean when we call BabelCoder an agentic framework.}

\begin{figure}[]
\centering
\begin{tcolorbox}[mypromptbox, width=\linewidth]
{\footnotesize
\textbf{\textcolor{purple}{``context''}}: You are an expert software developer and you can translate code from \textcolor{violet}{\{source\_lang\}} to \textcolor{violet}{\{target\_lang\}}.\\
\textbf{\textcolor{purple}{``prompt''}}: \textcolor{violet}{\{source\_code\}} \\
Translate the above \textcolor{violet}{\{source\_lang\}} code to \textcolor{violet}{\{target\_lang\}}. Print only the \textcolor{violet}{\{target\_lang\}} code and end with the comment $\backslash$End of Code$\backslash$. Do not give any other explanations or any other text except the \textcolor{violet}{\{target\_lang\}} code.\\
\textbf{\textcolor{purple}{``output''}}: \textcolor{violet}{\{translated\_code\}}
}
\end{tcolorbox}

\caption{Prompt template used by Translation Agent to translate code. The \{source\_code\} can be either the code in the source language or \NLSpec.}
\label{fig:prompt-translation}
\end{figure}

\subsection{Translation Agent}
\label{translation-agent}
The Translation Agent is the first agent activated after the user submits the source code. Its main role is to perform an initial code translation by querying a large language model (LLM). It constructs a prompt using the source code, or using the \NLSpec, where the \NLSpecS is obtained from the Refinement Agent. The constructed prompt follows a predefined template, shown in Figure~\ref{fig:prompt-translation}. After generating the translated code, the Translation Agent forwards it to the Refinement Agent for further processing and validation. Part A in Figure~\ref{fig:overview} represents the Translation Agent.

\subsection{Refinement Agent}
\label{refinement-agent}
The Refinement Agent improves the correctness of translated code that fails test cases through a fully automated, test-driven workflow. It collaborates with the Test Agent to analyze runtime feedback and with the Translation Agent when re-translation is needed. The agent automatically identifies, localizes, and fixes issues to ensure the code meets its intended functionality. It operates through the following workflows, illustrated in Part B of Figure~\ref{fig:overview}.

\subsubsection{Natural Language Specification (\NLSpec) Generation}
This step is invoked once at the beginning of the refinement workflow (Part B in Figure~\ref{fig:overview}). It takes the original source code as input and generates a Natural Language Specification (\NLSpec), which provides a line-by-line description of the program's logic and structure. The \NLSpec{} captures the semantic intent of each line while preserving the indentation and control flow of the original code. It omits language-specific syntax and focuses on expressing the core functionality in plain, language-agnostic terms. This specification helps the Translation Agent better interpret the source program’s intended behavior. The prompt template used for this generation is shown in Figure~\ref{fig:prompt-nlspec}.

\begin{figure}[]
\centering
\begin{tcolorbox}[mypromptbox, width=\linewidth]
{\footnotesize
\textbf{\textcolor{purple}{``prompt''}}: \textcolor{violet}{\{source\_code\}} \\
Give pseudocode for the above \textcolor{violet}{\{source\_language\}} code so that the \textcolor{violet}{\{source\_language\}} code is reproducible from the pseudocode. Do not give any other explanation except for the pseudocode. \\
\textbf{\textcolor{purple}{``output''}}: \textcolor{violet}{\{translated\_\NLSpec\}}
}
\end{tcolorbox}
\caption{Prompt template for generating \NLSpecS}
\label{fig:prompt-nlspec}
\end{figure}

\subsubsection{\NLSpecS Validation}
\label{nl-validation}

This step is responsible for validating and improving the generated \NLSpec. It takes the initial \NLSpecS as input and follows a multi-stage process to ensure the specification accurately reflects the intent of the source code. Once validated and refined, the updated \NLSpecS is passed to the next step. The key operations in this step are as follows:

\begin{itemize}
\item First, it analyzes the test report from the \textit{Test Agent} to identify failed test cases that may indicate incorrect or incomplete translations.
\item Next, it uses an LLM to perform line-by-line alignment between the source code and the \NLSpec. Based on this alignment, the LLM revises the \NLSpecS to better capture the behavior and logic of the original code.
\item Finally, a new translation is produced from the refined \NLSpecS using another LLM invocation.
\end{itemize}

Figure~\ref{fig:prompt-aligner} shows the prompt template used for aligning and refining the \NLSpec.

\begin{figure}[]
\centering
\begin{tcolorbox}[mypromptbox, width=\linewidth]
{\footnotesize
\textbf{\textcolor{purple}{``context''}}: You are an expert \textcolor{violet}{\{source\_lang\}} to \NLSpecS aligner. You will be given a \textcolor{violet}{\{source\_lang\}} code and corresponding \NLSpec. Your task is to align the \textcolor{violet}{\{source\_lang\}} code and the \NLSpecS line by line and update the \NLSpecS accordingly. Please return only the updated \NLSpecS without any further description.\\
\textbf{\textcolor{purple}{``prompt''}}: \textcolor{violet}{\{source\_lang\}}: \textcolor{violet}{\{source\_code\}} \\
Corresponding \NLSpecS: \textcolor{violet}{\{\NLSpec\}}\\
\textbf{\textcolor{purple}{``output''}}: \textcolor{violet}{\{updated\_\NLSpec\}}
}
\end{tcolorbox}
\caption{Prompt template for aligning \NLSpecS with translated code and refine the \NLSpecS using LLM.}
\label{fig:prompt-aligner}
\end{figure}

\subsubsection{\NLSpecS Augmentation}
\label{nl-agumentation}
Here, with the help of a state machine, the system decides whether to use {the original source code, the \NLSpecS, or both,} as input to the Translation Agent. The state machine governs this selection process by evaluating each input configuration to determine which leads to better translation outcomes. The evaluation is based on the number of passed test cases, with test execution handled by the Test Agent.
To further enhance the translation process, the system uses the updated \NLSpecS produced by the preceding \textit{\NLSpecS Validation} step. This refined \NLSpecS serves as one of the input options for the Translation Agent. The state machine tracks the current input configuration and updates it based on test case pass rates. During each refinement iteration, the Translation Agent generates candidate translations using the selected input form, and the state machine manages the coordination of input selection accordingly.

{To make this concrete, the state machine works over three input configurations, the source code alone, the validated \NLSpecS alone, and the source code and \NLSpecS together. The source code is used first. After each refinement cycle the machine records the number of tests the resulting translation passes, and if the current configuration does not improve on the best count seen so far it advances to the next configuration for the following cycle. The \NLSpecS configurations are therefore brought in when a translation still fails, and the chosen input form adapts to the observed test outcomes rather than being fixed in advance.}

\subsubsection{Fixing Errors}
\label{fixing-errors}
In this step, the refinement agent interacts with the test agent to verify and fix the translated code. The process is divided into two parts. First, the translated code is checked for compilation errors without generating test cases. If any compilation errors are found, they are fixed by invoking the LLM. After resolving compilation errors, if any other types of errors remain, they are addressed through another LLM invocation, using the summarized error message and the translated code.

\subsubsection{Bug Localization}
This workflow is responsible for identifying the most probable locations of bugs in the translated code. It leverages two complementary localization strategies to enhance accuracy and guide the repair process: (1) Spectrum-Based Fault Localization (SBFL) and (2) LLM-Based Scope Estimation. SBFL is effective when test cases and control-flow branches are available, as it highlights suspicious lines based on runtime behavior. In contrast, LLM-based localization identifies the likely scope of the bug by analyzing the structure and meaning of the code, especially when test feedback is unavailable or uninformative.

\paragraph{Spectrum-Based Fault Localization (SBFL).}
This component utilizes test execution data to compute a suspiciousness score for each line of code. It begins by collecting coverage information and test reports from the \textit{Test Agent}, using language-specific tools. Based on the pass/fail outcomes of the test cases, it constructs an SBFL matrix and ranks each line of code by its likelihood of containing a fault. The lines with the highest suspiciousness scores are selected as candidate bug locations. These locations are then passed to the next step, \textit{Fix error within scope}, to fix the bug.

\paragraph{LLM-Based Scope Estimation.}
\label{bug-scope-estimation}
To complement line-level localization, the LLM-based scope estimation employs an LLM to infer the likely scope of the bug. It categorizes the fault into one of several predefined code scopes: Input Processing,  Output Formatting, Variable Declaration, Loop Blocks, and Conditional Blocks. Using the original source code, the buggy translated code, and summarized error messages from the \textit{Test Agent}, the LLM analyzes the context and selects the most probable scope. It returns the scope category, estimated line ranges, and a natural language justification. This high-level guidance helps the system constrain the repair process to relevant sections of the code, making fixes more precise and efficient. The prompt template for finding the scope is shown in Figure~\ref{fig:prompt-scope}.

\begin{figure}[]
\centering
\begin{tcolorbox}[mypromptbox, width=\linewidth]
{\footnotesize
\textbf{\textcolor{purple}{``context''}}: You are an expert bug finder agent for \textcolor{violet}{\{type\}} in \textcolor{violet}{\{tgt\_lang\}} translated from \textcolor{violet}{\{src\_lang\}}. You will be provided a source code written in \textcolor{violet}{\{src\_lang\}}, a translated version of the code in \textcolor{violet}{\{tgt\_lang\}} that contains bugs, and the error message for the buggy code. Please check the following scopes and find out where the bug remains: \textcolor{violet}{\{scopes\}}. Return the scope of the bug with a precise and very brief explanation and line numbers.\\
\textbf{\textcolor{purple}{``prompt''}}: \textcolor{violet}{\{src\_lang}\} Source Code: \textcolor{violet}{\{src\_code}\} \\
Translated \textcolor{violet}{\{tgt\_lang}\} Buggy Code: \textcolor{violet}{\{trans\_code}\} \\
Error Message: \textcolor{violet}{\{error\_messages}\} \\
\textbf{\textcolor{purple}{``output''}}: \textcolor{violet}{\{Answer\}}
}
\end{tcolorbox}
\caption{Prompt template for finding scope in buggy translated code}
\label{fig:prompt-scope}
\end{figure}

\subsubsection{Fix Error within Scope} 
This step repairs the found errors using the provided information gathered from the previous steps. To fix the error, it invokes LLM with the translated buggy code and bug information, such as SBFL top suspicious lines, or bug scope information with line numbers. With the provided information, it sends the info to LLM and gets the repaired code. After this step, the \textit{\NLSpecS Augmentation} workflow is again used for the next cycle. The prompt used in this agent to fix the bug is shown in Figure~\ref{fig:prompt-fixscope}.

\begin{figure}[]
\centering
\begin{tcolorbox}[mypromptbox, width=\linewidth]
{\footnotesize
\textbf{\textcolor{purple}{``context''}}: You are an expert bug repair tool to solve \textcolor{violet}{\{type\}} in \textcolor{violet}{\{tgt\_lang\}} translated from \textcolor{violet}{\{src\_lang\}}. You will be provided with a code snippet in \textcolor{violet}{\{tgt\_lang\}} that contains bugs, and the possible bug location. Please fix the translated code using the bug description. Return only the fixed code without any additional text.\\
\textbf{\textcolor{purple}{``prompt''}}: \textcolor{violet}{\{src\_lang}\} Source Code: \textcolor{violet}{\{src\_code}\} \\
Translated \textcolor{violet}{\{tgt\_lang}\} Buggy Code: \textcolor{violet}{\{trans\_code}\} \\
Bug Location: \textcolor{violet}{\{scope}\} \\
\textbf{\textcolor{purple}{``output''}}: \textcolor{violet}{\{Answer\}}
}
\end{tcolorbox}
\caption{Prompt template to repair bug using LLM. \{type\} specifies the bug type (e.g., compilation error or assertion error).}
\label{fig:prompt-fixscope}
\end{figure}

\subsection{Test Agent}
\label{test-agent}
The Test Agent validates the translated code by executing generated test cases. 
It checks for several types of issues, including compilation errors, assertion failures, and runtime exceptions. The agent summarizes any encountered error messages and forwards them to the Refinement Agent to support subsequent repair steps. The steps followed by this agent are as follows. {When correctness is judged from function return values, the Test Agent also relies on a static unit test converter that turns Python-style assertions into executable JUnit tests, which we describe in Section~\ref{unit-test-conversion}.}

\subsubsection{Test Case Generation}
To validate the code provided by the Refinement Agent, this workflow generates a set of test cases using an LLM. {Each time the generator runs, it selects one test category uniformly at random from the three options in Figure~\ref{fig:prompt-testg}, namely \emph{complex}, \emph{difficult}, and \emph{corner case}, and requests from the model a batch of ten inputs in that category. The inputs are generated from the source program rather than the translated code, so that they reflect the intended behavior and are not shaped by any bug that translation may have introduced. Generation runs once in each refinement cycle. After every repair attempt, the updated translation is tested against a newly generated batch, so across the whole refinement process a program is usually exercised by inputs drawn from more than one category. We treat the three categories as a way to diversify the generated inputs rather than as a coverage target, and we do not require that every category appear for a given program.} These test cases are then used to evaluate the correctness of the translation. To prepare the expected outputs (oracles), the agent executes the original source code and records its output for comparison. {Both the generated inputs and their oracles therefore come from the source program, and the benchmark's ground-truth evaluation tests are never used during translation or repair. The signal that drives the internal repair loop is thus separate from the held-out suite on which we finally evaluate, so the reported accuracy reflects generalization rather than fitting to the evaluation tests.} {Coverage is measured separately. We collect it with language-specific tools and use it only inside the Spectrum-Based Fault Localization component (Section~\ref{bug-scope-estimation}) to rank suspicious lines, and we do not report it as a standalone accuracy metric.} The prompt template to generate the test cases using an LLM is shown in Figure~\ref{fig:prompt-testg}.
% { R1.3 The description of test case generation and coverage computation lacks clarity. While Fig.8 suggests that test cases are generated by randomly selecting among complex, difficult, and corner cases, Section 3.3.1 only mentions corner cases (Page 9 line 22), and it is unclear how many times the prompt is invoked or whether coverage across these categories is ensured. Clarifying the sampling strategy and its relationship to the reported coverage metrics would improve reproducibility.}

\begin{figure}[]
\centering
\begin{tcolorbox}[mypromptbox, width=\linewidth]

{
\footnotesize
\textbf{\textcolor{purple}{``context''}}: You are an expert Software Quality Assurance Engineer. You can write high quality and \textcolor{violet}{\{choice\}} tests for \textcolor{violet}{\{language\}} code. \\
\textbf{\textcolor{purple}{``prompt''}}: \textcolor{violet}{\{source\_code\}} \\
Generate \textcolor{violet}{\{no\_of\_tests\}} \textcolor{violet}{\{choice\}} input for the above \textcolor{violet}{\{language\}} code.\\
For your reference, a sample test case is as follows:\\
\textcolor{violet}{\{sample\_test\}}\\
Maintain the following output format (x will be \textcolor{violet}{\{no\_of\_tests - 1\}}):\\
Input\_0:\\
$<$input$>$\\
...\\
Input\_x:\\
$<$input$>$\\
Do not add any extra explanation or any other text except the mentioned output format.
\\
\textbf{\textcolor{purple}{``output''}}: \textcolor{violet}{\{Answer\}}

}
\end{tcolorbox}
\caption{Prompt template for generating test cases. (\textcolor{violet}{choice} is randomly selected from the three: complex, difficult, corner case.)}
\label{fig:prompt-testg}
\end{figure}

\subsubsection{Program Execution}
This step is for executing the translated target code using the test cases generated by the Test Case Generation workflow. It detects error types that occur during compilation or runtime. Upon encountering errors, it extracts and summarizes the error messages by filtering out only the relevant parts and removing repetitive or non-informative elements such as file paths, hexadecimal values, and other unnecessary details. After summarizing, it generates a report for each test case execution.

\subsubsection{Generate Coverage and Test Reports}
After executing the target code with the generated test cases, the system produces a test report along with a coverage report. The coverage report is later utilized by the refinement agent. Language-specific coverage tools are employed to generate coverage metrics across different programming languages.

\section{Experiment Setup}
\subsection{Datasets}
\label{datasets}
We use four widely adopted datasets for evaluating code translation and generation, covering a diverse set of programming languages (C, C++, Go, Java, Python). 
 
\pa{Avatar~\citep{ahmad-etal-2021-avatar}}. This dataset contains 9,515 code snippets written in Python and Java, collected from online competitive programming platforms. It includes test cases for 250 problems and provides aligned function-level code across languages. We consider all possible translation directions from two source languages (Python and Java) to five target languages (Python, Java, C, C++, and Go), resulting in eight unique language pairs. In this work, we use the subset curated by Pan et al.~\citep{pan2024lost}, which selected only the samples that include test cases (250 problems per language). We filter out samples where the ground-truth solutions fail the provided tests, resulting in 240 valid samples in Java and 239 in Python, yielding a total of {479 data points and 9,381 test cases (4,691 for Java and 4,690 for Python)}.

\pa{CodeNet~\citep{puri2021codenet}}. A large-scale dataset collected from online programming competitions, containing 14 million code samples across 55 programming languages. In this work, we focus on five languages (C, C++, Go, Java, and Python) as both source and target languages, resulting in 20 language pairs. We directly use the subset curated by Pan et al.~\citep{pan2024lost} without any modification, which includes 200 valid instances per language, totaling 1,000 code samples. Each instance is accompanied by a single test case. 

\pa{EvalPlus~\citep{liu2024your}}. EvalPlus is an enhanced version of the handcrafted HumanEval dataset~\citep{chen2021evaluating}, written in Python with 164 samples. It improves the ground truth implementations and provides stronger, more comprehensive test cases. In this work, we evaluate the Python-to-Java translation task using EvalPlus test cases. Since EvalPlus only contains Python implementations, we use Java ground truths curated by Rabbi et al.~\citep{rabbi2025multi}, which are based on HumanEval-X~\citep{zheng2023codegeex} and aligned with EvalPlus test cases.

\pa{TransCoder~\citep{lachaux2020unsupervised}}. The code samples in this dataset are collected from GeeksForGeeks in three languages: C++, Java, and Python. In this work, we use the cleaned version released by Yang et al.~\citep{yang2024exploring}, which contains 464 standalone functions. We focus on the Python-to-Java direction, as our unit test converter currently supports only this translation pair. We obtained the original TransCoder dataset from previous work~\citep{lachaux2020unsupervised} and selected the same 464 samples as used in UniTrans~\citep{yang2024exploring} (one of our baselines) to enable direct comparison. Since the dataset does not include test cases, we automatically generated test cases from the provided Python code. Among the samples, 33 functions either produce standard output (e.g., \texttt{print}) or perform in-place modifications on input parameters without returning values, making them unsuitable for return-value-based evaluation. This issue was not addressed in previous works using this dataset, such as UniTrans~\citep{yang2024exploring} and TransAgent~\citep{yuan2024transagentllmbasedmultiagentcode}. To support meaningful evaluation, we manually added appropriate return statements to these functions. We include these modified datapoints in our artifact to support reproducibility and future research.

Table~\ref{tab:dataset} summarizes the statistics of the four datasets.

\begin{table}[]
\centering
\small
\renewcommand{\arraystretch}{1.2}
\caption{Statistics of the evaluation datasets. Note that TransCoder does not have ground truth test cases and relies on a test agent to generate test cases.}
% \resizebox{\textwidth}{!}{
% \begin{tabular}{|l|r|r|r|r|}
\begin{tabular}{p{2.1cm}|p{1.5cm}|p{1.5cm}|p{1.5cm}}
\hline

\textbf{Dataset} & \textbf{Source Language} & \textbf{Datapoints} & \textbf{Total \# Test Cases}\\
\midrule

\multirow{2}{*}{Avatar \citep{ahmad-etal-2021-avatar}} & Java & 240 & 4691 \\
& Python & 239 & 4690 \\
\midrule
\multirow{5}{*}{CodeNet \citep{puri2021codenet}} & C & 200 & 200\\
& C++ & 200 & 200\\
& Go & 200 & 200\\
& Java & 200 & 200\\
& Python & 200 & 200\\
\midrule
EvalPlus \citep{liu2024your} & Python & 164 & 2681\\
\midrule
TransCoder \citep{lachaux2020unsupervised} & Python & 464 & -\\
% \hline
\end{tabular}
% }
\label{tab:dataset}
\end{table}

\subsection{Unit Test Conversion}
\label{unit-test-conversion}
{For the benchmarks that we evaluate by functional correctness, namely EvalPlus and TransCoder, the Test Agent judges a translation by comparing the return value of the translated function against an expected value. EvalPlus ships its ground-truth oracles as Python assertions in the PyTest style, while TransCoder provides no test cases, so for TransCoder the Test Agent generates inputs and obtains the expected values by executing the source program. In both cases the oracle is a Python-side assertion that has to be compiled and executed in the target language, so we implement a static unit test converter that turns each such assertion into an equivalent JUnit test. The converter parses the assertion with Python's abstract syntax tree and reads the literal argument values and the expected output. From these values it infers the Java parameter and return types, covering scalars as well as nested and mixed element lists such as \texttt{list[int]}, \texttt{list[float]}, and \texttt{list[list[float]]}, and it rewrites the Python literals into their Java counterparts. During this step it respects the numeric ranges of the Java types, for example the limits of \texttt{Integer}, \texttt{Long}, \texttt{Float}, and \texttt{Double}, so that a value that would overflow a narrower type is assigned a wider one. It then reconstructs the JUnit test header from the inferred signature. The converter produces the JUnit tests that the Test Agent executes against the translated Java code, both for the tests generated during refinement and for the ground truth tests used in the final EvalPlus evaluation. Because its type inference and literal rewriting are implemented for the Python to Java direction, our functional evaluation uses that direction.}

{Avatar and CodeNet are different: they provide whole programs that read from standard input and print to standard output. The Test Agent checks these by comparing the printed output against the expected output, which does not depend on the target language, so no converter is needed. This is why EvalPlus and TransCoder are limited to Python-to-Java, while Avatar and CodeNet are evaluated across all of their language pairs. Extending EvalPlus or TransCoder to another target language would require adding the corresponding type and literal rules to the converter, which we leave for future work.}

\subsection{Studied LLMs and Hyper-Parameter Settings}
For a fair comparison with LIT work~\citep{pan2024lost} and TransAgent~\citep{yuan2024transagentllmbasedmultiagentcode}, we chose the GPT-4o model as the underlying LLM for \tool{} on the datasets of Avatar, CodeNet, EvalPlus, and TransCoder. 
GPT-4o is therefore the primary model for all reported results. The one exception is the TransCoder comparison with UniTrans~\citep{yang2024exploring}, which reports GPT-3.5 results, so we additionally run \tool{} with GPT-3.5 on that dataset to match their setting; both configurations are reported separately in Table~\ref{tab:translation_accuracy}.

%GPT-4o is a multimodal model developed by OpenAI that supports text, vision, and audio inputs. In this work, we use only its text capabilities via the OpenAI API. More implementation details are provided below.

%\subsection{Prompting Strategies}
For all experiments, we use a temperature setting of 0.8. While GPT-4o’s default temperature is 1.0 (encouraging diverse outputs), a lower value like 0.0 produces deterministic responses, and values approaching 2.0 introduce high randomness. Through experimentation, we found that a temperature of 0.8 balances diversity and relevance well, producing a useful range of candidate outputs. We set the maximum token limit to 8000 and customized the context for each agent accordingly.

\subsection{Baselines}
\label{baselines}
We compare our approach against four baselines: LIT~\citep{pan2024lost}, UniTrans~\citep{yang2024exploring}, TransAgent \citep{yuan2024transagentllmbasedmultiagentcode}, and InterTrans \citep{macedo2024intertrans}. 

\noindent\textbf{LIT.} The first is Lost in Translation \citep{pan2024lost}, which evaluates multilingual code translation using AVATAR, CodeNet, and EvalPlus datasets. They also included two real world projects Apache Click and Common CLI in their experiment each containing 16 datapoints. But according to Pan et al. \citep{pan2024lost} all the translations for these two datasets failed. Also, as no unit tests were available with real world datasets, we excluded them from our experiment. 

\noindent\textbf{UniTrans.} The second baseline, UniTrans \citep{yang2024exploring}, leverages the dataset introduced by \citep{lachaux2020unsupervised}.
This work shares similarities with LIT, as both propose autonomous translation frameworks combined with iterative repair mechanisms. However, the main difference is the integration of test case generation within the framework to assess computational accuracy (CA).

\noindent\textbf{TransAgent.} TransAgent \citep{yuan2024transagentllmbasedmultiagentcode} uses the same dataset as UniTrans and also uses UniTrans as the initial translation agent with a novel Alignment Agent which aligns source and translated code block by block (i.e., for, if, etc.) and uses the internal variables' values to find inconsistencies between the source and translated code.

\noindent\textbf{InterTrans.} 
\label{intertrans-baseline}
InterTrans~\citep{macedo2024intertrans} conducted experiments using three datasets: CodeNet~\citep{puri2021codenet}, HumanEval-X~\citep{chen2021evaluating}, and the dataset by Lachaux et al.~\citep{lachaux2020unsupervised}. Their best performance (95.4\%) was reported on HumanEval-X. { Unlike \tool, InterTrans does not use a GPT model. It evaluates open models such as Magicoder and StarCoder2, and Magicoder gives its best results. It also places the target function's method signature in the translation prompt, which makes the translation easier. Because InterTrans differs from \tool{} in its model, its benchmark, and its prompt, a comparison under a shared model is not possible, so we compare on the code that InterTrans released rather than rerunning it. We take this route because rerunning InterTrans under GPT-4o would replace the configuration its authors tuned and reported, so any drop in its accuracy would reflect our setup rather than the method itself. Comparing on the released outputs keeps InterTrans in the setting where it performs best, and we score both systems on the same 31 instances with the same EvalPlus tests, so the difference reflects the translations and not the evaluation.} However, as their evaluation was conducted on only 35 samples per language and dataset, making a direct comparison was challenging. To address this, we retrieved their generated code from the provided replication package, focusing specifically on the Python-to-Java translation results from HumanEval-X. Out of 35 samples, we exclude 4 as these HumanEval ground truths fail on EvalPlus test cases. We then compare the results of InterTrans based on these 31 samples with \tool.

{ We do not adopt Magicoder inside \tool, because unlike GPT-4o it does not reliably follow the varied instructions that our auxiliary agents need for test generation, \NLSpecS alignment, and bug scope classification.}

For comparing the baselines' results with ours, we reported the results directly from the baseline papers. 

\subsection{Evaluation Metrics}

\noindent We use computational accuracy (CA) as the primary evaluation metric across all experiments, following prior works \citep{lachaux2020unsupervised}, \citep{roziere2021leveraging}, \citep{szafraniec2022code}, \citep{liu2023syntax}, \citep{yang2024exploring}, \citep{pan2024lost}. CA indicates the ratio of translated codes that produce similar execution results or function outputs when given the same inputs as the source code, relative to the total number of translations. For Avatar and CodeNet, which provide complete programs that read from standard input and print output, we evaluate correctness by executing the program with given inputs and comparing the printed output to the ground truth. For EvalPlus and TransCoder, which involve generating or translating individual functions, we evaluate correctness by executing the generated function and directly comparing its return values to the expected outputs. 
Specifically, we assess correctness based on the exact match in case of $Strings$ and rounded value to three decimal points in case of $Integers, Doubles, Floats$ datatypes. 

\subsection{Cost of Experiment}

\noindent Across all four datasets, the experimental pipeline processed 5{,}744 datapoints, incurring a total cost of \$244.94~USD, which corresponds to an average of \$0.0453~USD per correct translation. {This total covers the main end-to-end evaluation of Section~\ref{results} together with the four broad ablation stages of Section~\ref{rq3}, and is therefore not the cost of a single translation. The reported total excludes the five-run stability study and the seven-configuration detailed ablation, which were executed separately. A single translation issues about $15$ LLM calls on the nominal path and at most $29$, which we break down by stage in the LLM Usage analysis of RQ1.}

{A direct comparison of monetary cost across methods is not possible, because none of the baselines report their dollar cost. We therefore compare on the number of LLM invocations per translation, a price-independent proxy for cost. We compare invocations rather than tokens because none of the baselines report token usage, and InterTrans's released artifacts record translation attempts rather than prompt sizes. We note that invocation counts do not capture per-call prompt size, and \tool{}'s later-stage prompts carry additional context such as the \NLSpecS{} and summarized error messages. As noted above, \tool{} issues about $15$ calls per instance when it traverses all four stages, as few as $2$ when the first translation compiles and passes its tests, and at most $29$. Among the baselines, only InterTrans releases enough run data to recover an invocation count. Because it searches a tree of transitive translation paths and stops at the first path that passes the tests, its cost depends on how many paths it must try: on its released HumanEval-X runs it averages about one call for pairs where the direct translation already passes (such as those targeting JavaScript or Python) and between $24.9$ and $30.9$ calls for pairs that translate Python or Java into C++, Go, or Rust, with a reported ceiling of up to $313$ calls for a single translation. Its average is $11.8$ calls across all $30$ pairs of that benchmark and $16.1$ on Python$\rightarrow$Java, measured over the $35$ instances it released for that pair. \tool{} instead issues a fixed, bounded number of calls per translation, whatever the language pair. RQ1 breaks our count down by stage and discusses this comparison in more detail.}
\section{Results}
\label{results}

\subsection*{RQ1: \RQOne}
\pa{Motivation.} Existing works evaluate code translation using different benchmarks, CodeNet and Avatar using standard input/output, and TransCoder and EvalPlus using functional correctness. To enable a comprehensive comparison, RQ1 investigates whether our agent-based framework can consistently outperform these approaches across all four benchmarks. In addition, because \tool{} operates at a non-zero temperature and is therefore non-deterministic, we also assess whether its accuracy is stable across repeated runs.

\pa{Method.} To answer this question, we run \tool{} end-to-end on the datasets in Section~\ref{datasets}. Then, we compare the results against the baselines outlined in Section~\ref{baselines}. For Avatar and CodeNet, we use the same samples as LIT~\citep{pan2024lost}, ensuring a fair comparison. For EvalPlus, we conduct experiments on the full dataset. Since UniTrans~\citep{yang2024exploring} uses GPT-3.5 on the TransCoder~\citep{lachaux2020unsupervised} dataset, we also use GPT-3.5 in our experiments for fair comparison. Similarly, as TransAgent uses GPT-4o on the TransCoder dataset, we used the same model in our experiments.

\begin{table*}[]
\small
\centering
\renewcommand{\arraystretch}{1.1}
\caption{Computational accuracy (\%) of BabelCoder across various source and target languages. (81.33\% CA in the UniTrans artifact might have been overestimated due to overlooking functions with missing return statements, and cannot be verified due to missing translated outputs in the artifact. Details in Section \ref{datasets}). LIT: Lost In Translation, UT: UniTrans, TA: TransAgent, IT: InterTrans, gen: generated test cases, eval: evaluation test cases provided by the dataset. {The (gen., \%) column reports \tool's accuracy on the tests generated by the Test Agent, and the (eval., \%) column reports its accuracy on the ground-truth evaluation tests shipped with the benchmark. The former is the signal that the internal repair loop optimizes, and the latter is the fixed held-out suite used for a fair comparison with the baselines. A dash in a cell indicates that the corresponding baseline did not report a result for that source and target pair, since the baselines cover different subsets of the datasets and language pairs.}}
\begin{tabular}{|p{2.1cm}|p{1.0cm}|p{1.0cm}|p{1.5cm}|p{1.5cm}|p{.7cm}|p{.7cm}|p{.7cm}|p{.7cm}|}

\hline
\textbf{Dataset} & \textbf{Source} & \textbf{Target} & \textbf{BabelCoder} & \textbf{BabelCoder} & \textbf{LIT} & \textbf{UT} & \textbf{TA} & \textbf{IT} \\
\textbf{} & \textbf{} & \textbf{} & \textbf{(gen., \%)} & \textbf{(eval., \%)} & \textbf{~\citep{pan2024lost}} & \textbf{~\citep{yang2024exploring}} & \textbf{~\citep{yuan2024transagentllmbasedmultiagentcode}} & \textbf{~\citep{macedo2024intertrans}} \\
\hline
\multirow{8}{*}{Avatar} & \multirow{4}{*}{Java}     & C      & 97.92 & \cellcolor{gray!30} \textbf{95.83} & 82.33 & --    & -- & -- \\
                        &                           & C++    & 97.50 & \cellcolor{gray!30} \textbf{96.25} & 91.97 & --    & -- & -- \\
                        &                           & Go     & 90.83 & \cellcolor{gray!30} \textbf{87.50} & 80.72 & --    & -- & -- \\
                        &                           & Python & 94.58 & \cellcolor{gray!30} \textbf{92.50} & 85.94 & --    & -- & -- \\ \cline{2-9}
                        & \multirow{4}{*}{Python}   & C      & 91.67 & \cellcolor{gray!30} \textbf{86.67} & 62.00 & --    & -- & -- \\
                        &                           & C++    & 94.17 & \cellcolor{gray!30} \textbf{87.92} & 70.00 & --    & -- & -- \\
                        &                           & Go     & 86.67 & \cellcolor{gray!30} \textbf{84.17} & 63.60 & --    & -- & -- \\
                        &                           & Java   & 90.83 & \cellcolor{gray!30} \textbf{88.33} & 76.40 & --    & -- & -- \\
\hline
\multirow{20}{*}{CodeNet}   & \multirow{4}{*}{C}        & C++    & 94.50 & \cellcolor{gray!30} \textbf{99.00} & 98.50 & --    & -- & -- \\
                            &                           & Go     & 79.50 & 87.50 & \cellcolor{gray!30} \textbf{93.00} & --    & -- & -- \\
                            &                           & Java   & 91.50 & \cellcolor{gray!30} \textbf{98.00} & 96.50 & --    & -- & -- \\
                            &                           & Python & 84.50 & \cellcolor{gray!30} \textbf{93.00} & 89.50 & --    & -- & -- \\ \cline{2-9}
                            & \multirow{4}{*}{C++}      & C      & 95.00 & \cellcolor{gray!30} \textbf{98.00} & 97.00 & --    & -- & -- \\
                            &                           & Go     & 85.50 & 94.00 & \cellcolor{gray!30} \textbf{94.50} & --    & -- & -- \\
                            &                           & Java   & 93.50 & \cellcolor{gray!30} \textbf{98.00} & 97.50 & --    & -- & -- \\
                            &                           & Python & 93.00 & \cellcolor{gray!30} \textbf{94.00} & 89.50 & --    & -- & -- \\ \cline{2-9}
                            & \multirow{4}{*}{Go}       & C      & 95.00 & \cellcolor{gray!30}\textbf{97.50} & 96.00 & --    & -- & -- \\
                            &                           & C++    & 98.00 & \cellcolor{gray!30}\textbf{98.50} & 97.00 & --    & -- & -- \\
                            &                           & Java   & 95.50 & \cellcolor{gray!30} \textbf{99.00} & 95.50 & --    & -- & -- \\
                            &                           & Python & 94.00 & \cellcolor{gray!30} \textbf{96.50} & 90.50 & --    & -- & -- \\ \cline{2-9}
                            & \multirow{4}{*}{Java}     & C      & 92.50 & \cellcolor{gray!30} \textbf{97.50} & 93.50 & --    & -- & -- \\
                            &                           & C++    & 92.00 & \cellcolor{gray!30} \textbf{98.00} & 96.00 & --    & -- & -- \\
                            &                           & Go     & 77.00 & 87.50 & \cellcolor{gray!30} \textbf{91.50} & --    & -- & -- \\
                            &                           & Python & 88.50 & \cellcolor{gray!30} \textbf{94.00} & 91.50 & --    & -- & -- \\ \cline{2-9}
                            & \multirow{4}{*}{Python}   & C      & 94.50 & \cellcolor{gray!30} \textbf{98.00} & 94.50 & --    & -- & -- \\
                            &                           & C++    & 97.00 & \cellcolor{gray!30} \textbf{99.50} & 96.00 & --    & -- & -- \\
                            &                           & Go     & 90.00 & \cellcolor{gray!30} \textbf{94.00} & 88.00 & --    & -- & -- \\
                            &                           & Java   & 96.00 & \cellcolor{gray!30} \textbf{99.00} & 96.50 & --    & -- & -- \\
\hline
EvalPlus                 & Python & Java & 93.29 & \cellcolor{gray!30} \textbf{91.46} & 84.76 & 80.49 & -- & -- \\
\hline
EvalPlus          (31 samples)                 & Python & Java & -- & \cellcolor{gray!30} \textbf{87.01} & -- & -- & -- & \cellcolor{gray!30} \textbf{87.01} \\
\hline
TransCoder (GPT-3.5)               & Python & Java & \cellcolor{gray!30} \textbf{90.52} & -- & -- & 81.33 & -- & -- \\
\hline
TransCoder (GPT-4o)              & Python & Java & \cellcolor{gray!30} \textbf{92.24} & -- & -- & -- & 89.50 & -- \\
\hline
\end{tabular}
\label{tab:translation_accuracy}
\end{table*}

\pa{Result.} The outcomes are summarized in Table~\ref{tab:translation_accuracy}. For \tool, we reported the CA for both provided test cases for comparison with LIT and generated test cases for comparison with UniTrans and TransAgent.
As seen in the table, our framework consistently outperforms the baselines on 26 of the 29 language pairs under the same setting, with gains of up to 24.67 percentage points (pp). The improvements are largest where the source and target languages differ most in syntax and idiom, reaching +24.67\,pp on Avatar Python$\rightarrow$C, +20.57\,pp on Python$\rightarrow$Go, and +17.92\,pp on Python$\rightarrow$C++. The three remaining pairs are all Go targets in CodeNet. On C++-Go the two systems are effectively tied, differing by 0.50\,pp, which is a single instance out of 200, while on C-Go and Java-Go LIT is ahead by 5.50\,pp and 4.00\,pp (11 and 8 instances). In these three cases LIT is ahead of \tool{}, by 5.50\,pp on C-Go, 4.00\,pp on Java-Go, and 0.50\,pp on C++-Go.
We believe this discrepancy stems from the nature of the CodeNet evaluation suite, which provides only a single test case per sample. This tends to produce more conservative or generic translations, which may happen to pass the single evaluation test more often, even if they are less correct overall. In contrast, our framework relies on multiple GPT-4o-generated tests to guide the translation and repair process. The limited evaluation coverage (single tests) may not fully reflect the improvements made by our method.

Overall, results using the evaluation tests are consistently better than those using GPT-4o-generated tests, which indicates that GPT-4o is capable of generating diverse and more challenging test cases that facilitate effective program repair.

As described in Section~\ref{intertrans-baseline}, we compare \tool{} with InterTrans using the same 31 samples. InterTrans reaches the same accuracy on these samples but does so through a variable-cost search over transitive translation paths and by placing the target function's header in its prompt, whereas \tool{} matches it with a fixed, bounded pipeline and without the header. We quantify the invocation cost below (LLM Usage).

{\noindent \textbf{Run-to-run stability.} Because \tool{} operates at a non-zero decoding temperature ($T = 0.8$), its multi-agent pipeline is non-deterministic: it chains up to 29 LLM calls per instance at worst case, where each agent conditions on the output of the previous one. To verify that the strong performance reported above is stable rather than the product of a single favorable run, we execute the full pipeline five independent times on EvalPlus (Python$\rightarrow$Java), keeping all hyperparameters fixed. Each run is a complete, independent end-to-end execution with fresh intermediate states. Since the traditional pass@$k$ metric~\citep{chen2021evaluating} is designed for single-call generation, where $k$ samples are drawn independently from one LLM invocation, it is architecturally inappropriate for a stateful, sequential pipeline such as \tool{}. We therefore adapt it to the pipeline level: for each of the 164 instances we count the runs $c_i$ (out of $n=5$) that yield a correct translation, and compute pass@$k = \mathbb{E}_i[1 - \binom{n-c_i}{k}/\binom{n}{k}]$, the expected probability that at least one of $k$ independent pipeline runs succeeds. As reported in Table~\ref{tab:temperature_sensitivity}, \tool{} achieves a mean CA of $90.58 \pm 0.72$\% across the five runs, with individual runs ranging narrowly from 89.6\% to 91.5\%. This low standard deviation confirms that the pipeline produces stable and consistent results despite its non-deterministic temperature, while the pass@5 of 93.3\%, higher than any single run, reflects that independent re-execution reliably recovers translations that occasionally fail in a given run. Together, these results demonstrate that the non-determinism introduced by $T = 0.8$ does not undermine the reliability of \tool{}.}

\begin{table}
\centering
\caption{Computational Accuracy (\%) of \tool{} across 5 independent runs on EvalPlus (Python source, 164 instances, GPT-4o, temperature = 0.8). $\mu \pm \sigma$ = mean $\pm$ standard deviation.}
\label{tab:temperature_sensitivity}
\resizebox{\columnwidth}{!}{%
\begin{tabular}{l|l|ccccc|cc}
\toprule
\textbf{Dataset} & \textbf{Lang. Pair} & \textbf{Run 1} & \textbf{Run 2} & \textbf{Run 3} & \textbf{Run 4} & \textbf{Run 5} & $\boldsymbol{\mu \pm \sigma}$ & \textbf{pass@5} \\
\midrule
EvalPlus & Python $\rightarrow$ Java   & 90.8 & 90.2 & 90.8 & 91.5 & 89.6 & $90.58 \pm 0.72$ & 93.3 \\
\bottomrule
\end{tabular}%
}
\end{table}

{ \noindent \textbf{LLM Usage.} We quantify efficiency as the number of LLM
invocations per translated instance. \tool{} runs four stages, and each stage issues a fixed set of calls whose repair loops are capped at three attempts. Stage~(i) performs the initial translation and issues one translation call, one test-generation call, and up to three calls for compilation repair and three for runtime repair, so it costs four calls on the nominal path and at most eight. Stage~(ii) generates the \NLSpecS{} and re-translates, adding one \NLSpecS{} generation call, one translation call, one test-generation call, and the same two repair loops, for five calls nominally and at most nine. Stage~(iii) aligns and validates the \NLSpecS{} and re-translates with the same structure, again five calls nominally and at most nine. Stage~(iv) classifies the bug scope and repairs within it, for one call nominally and at most three. A translation therefore costs about $15$ calls when it traverses all four stages and every repair succeeds on its first attempt, and is bounded above by $29$ calls when every repair loop is exhausted. The minimum is $2$ calls: an instance whose translation compiles and passes its generated tests returns at the end of Stage~(i), having issued only one translation call and one test-generation call. This ceiling is fixed by the pipeline and does not grow with the difficulty of the language pair or the size of the program. The key property is that this budget is fixed in advance: it depends only on how many repairs are triggered, never on the language pair or the size of the program. InterTrans has no such fixed cost. It searches a tree of transitive translation paths and stops at the first path that passes the tests, so its number of LLM calls depends on how many paths it must try before one succeeds. We measured this on InterTrans's released HumanEval-X runs with Magicoder, its strongest model. When the direct one-step translation already passes, InterTrans stops at a single call, which is why pairs that target JavaScript or Python average close to one call (for example, $1.2$ for C++$\rightarrow$JavaScript). When the direct translation usually fails, it searches longer, so pairs that translate Python or Java into C++, Go, or Rust average between $24.9$ and $30.9$ calls, and its authors report a ceiling of up to $313$ calls for a single translation. On Python$\rightarrow$Java, the pair on which we compare accuracy in Table~\ref{tab:translation_accuracy}, InterTrans issues $16.1$ calls per translation across the $35$ HumanEval-X instances it released for that pair. Averaged over all $30$ language pairs of its benchmark, including pairs involving JavaScript and Rust that we do not evaluate, it issues $11.8$. Neither average is a reliable guide to a given instance, because the cost rises with the number of failing paths and has no bound fixed in advance. \tool{} instead stays at about $15$ calls and never exceeds $29$, whatever the language pair or program size. On the shared 31-instance subset (Section~\ref{intertrans-baseline}), \tool{} reaches the same accuracy as InterTrans with this fixed, predictable budget rather than an open-ended search.}

% \begin{table}
% \centering
% \caption{{LLM invocations per translated instance. Values marked \emph{reported}
% are taken from the corresponding paper, and those marked \emph{derived} are obtained by
% counting the invocation sites in the pipeline the paper describes. A dash indicates that no
% fixed ceiling is stated.}}
% \label{tab:llm_usage}
% \begin{tabular}{llll}
% \toprule
% \textbf{Approach} & \textbf{Repair loop} & \textbf{Calls per instance} & \textbf{Basis} \\
% \midrule
% LIT~\citep{pan2024lost}                                  & iterative, no cap & not reported & -- \\
% UniTrans~\citep{yang2024exploring}                        & $3$ rounds        & $\approx 5$  & reported, derived \\
% TransAGENT~\citep{yuan2024transagentllmbasedmultiagentcode} & dynamic         & not bounded  & reported \\
% InterTrans~\citep{macedo2024intertrans}                   & --                & up to $313$  & reported \\
% \tool{}                                                   & $3$ per stage     & $\approx 15$, at most $29$ & derived \\
% \bottomrule
% \end{tabular}
% \end{table}

\subsection*{RQ2. \RQTwo}
\label{rq2}
\pa{Motivation.} During code translation between significantly different programming languages, software quality, especially security, becomes a major concern. This research question investigates the extent to which translation affects code quality, focusing specifically on security-related issues and general software defects before and after translation.

\pa{Method.} To address this question, we analyze both the original source code and the corresponding translated code from our four datasets using \textit{SonarQube}~\citep{sonarqube}. {The translated code in this analysis is the output of \tool{} itself, produced with GPT-4o, and not the output of any baseline. The comparison therefore measures how our own translations affect code quality relative to the original source.} SonarQube is chosen for its ability to automatically detect a wide range of software quality issues, including security vulnerabilities (e.g., injection flaws, access control) and maintainability issues (e.g., code smells, duplications). Our analysis concentrates on two key categories reported by \textit{SonarQube}: blocker and critical issues, which typically include severe bugs and security vulnerabilities. In addition, we consider any flagged security hotspots or direct vulnerabilities identified in the code. To normalize the comparison across files of varying size, we report the number of issues per 1,000 lines of non-commented source code (NLOC). 

{Two clarifications are in order. First, RQ2 measures a downstream effect of translation rather than an agentic mechanism. Whether moving code from one language to another introduces quality or security issues matters directly to anyone who deploys an automatic translator such as \tool, which is why we study it. Second, we do not treat SonarQube as ground truth. Static analyzers produce false positives, and their false-positive rates differ across languages, so we do not assume that this bias cancels in a cross-language comparison. For this reason we analyze the source and the translated code under an identical configuration, we report normalized before and after rates and interpret their differences rather than treating raw counts as ground truth, and we restrict attention to blocker and critical issues and to flagged security hotspots, which are the categories least affected by stylistic false positives. We therefore read the SonarQube results as an indicative signal of how translation shifts code quality, not as a definitive security measurement.}

\pa{Result.} Table~\ref{tab:translation_issues} presents the identified issues and associated security impacts in both the source and translated code across all selected datasets. The results show that when translating from other languages to C, the number of issues consistently increases by 18.35 to 35.86 per 1000 NLOC. A similar trend is observed for security impacts, which also rise by 4.9 to 10.82 per 1000 NLOC when translated into C. One possible explanation for this behavior is that C lacks many high-level abstractions and safety features (e.g., memory management, type safety) available in other languages, making it more prone to translation-related bugs and vulnerabilities.
Interestingly, when translating from C++ to C, the number of issues does not increase. In contrast, translating from C to C++ produces 18.66 more issues per 1000 NLOC. This observation may offer a useful insight: to reduce translation-induced issues, it could be beneficial to use an intermediate language such as C++. For example, translating from another language to C++ and then to C might help limit the number of introduced issues. A deeper investigation into this approach is beyond the scope of this study and is left for future work.

\begin{table}[]
\small
\centering
\caption{Number of Blocker \& Critical issues and Security Impacts before and after code translation (per 1K NLOC).}
\begin{tabular}{|p{1.2cm}|p{1cm}|p{1cm}|p{1cm}|p{1cm}|p{1cm}|p{1cm}|}
\hline
\multirow{2}{*}{\textbf{Dataset}} & \multirow{2}{*}{\textbf{Source}} & \multirow{2}{*}{\textbf{Target}} & \multicolumn{2}{|c|}{\textbf{Issues}} & \multicolumn{2}{|c|}{\textbf{Security Impacts}} \\
\cline{4-7}
 &  &  & \textbf{Before} & \textbf{After} & \textbf{Before} & \textbf{After} \\
\hline
\multirow{8}{*}{Avatar} & \multirow{4}{*}{Java}     & C      & \multirow{4}{*}{16.30} & \cellcolor{gray!30} \textbf{34.65} & \multirow{4}{*}{0} & \cellcolor{gray!30} \textbf{6.28} \\
                        &                           & C++    &  & \cellcolor{gray!30} \textbf{19.0}  &  & 0 \\
                        &                           & Go     &  & 4.46  &  & 0 \\
                        &                           & Python &  & 6.04  &  & 0 \\ \cline{2-7}
                        & \multirow{4}{*}{Python}   & C      & \multirow{4}{*}{15.03} & \cellcolor{gray!30} \textbf{39.54} & \multirow{4}{*}{0} & \cellcolor{gray!30} \textbf{6.27} \\
                        &                           & C++    &  & 2.85  &  & 0 \\
                        &                           & Go     &  & 2.19  &  & 0 \\
                        &                           & Java   &  & 0.51  &  & 0 \\
\cline{1-7}
\multirow{20}{*}{CodeNet}   & \multirow{4}{*}{C}        & C++    & \multirow{4}{*}{41.28} & \cellcolor{gray!30} \textbf{59.94} & \multirow{4}{*}{5.41} & 0.15 \\
                            &                           & Go     &  & 3.47  &  & 0 \\
                            &                           & Java   &  & 6.58  &  & 0 \\
                            &                           & Python &  & 4.34  &  & 0 \\ \cline{2-7}
                            & \multirow{4}{*}{C++}      & C      & \multirow{4}{*}{51.27} & 38.12 & \multirow{4}{*}{1.03} & \cellcolor{gray!30} \textbf{5.93} \\
                            &                           & Go     &  & 3.35  &  & 0 \\
                            &                           & Java   &  & 7.15  &  & 0 \\
                            &                           & Python &  & 3.92  &  & 0 \\ \cline{2-7}
                            & \multirow{4}{*}{Go}       & C      & \multirow{4}{*}{1.84} & \cellcolor{gray!30} \textbf{37.07} & \multirow{4}{*}{0} & \cellcolor{gray!30} \textbf{5.57} \\
                            &                           & C++    &  & \cellcolor{gray!30} \textbf{16.4}  &  & 0 \\
                            &                           & Java   &  & \cellcolor{gray!30} \textbf{7.59}  &  & 0 \\
                            &                           & Python &  & \cellcolor{gray!30} \textbf{4.79}  &  & 0 \\ \cline{2-7}
                            & \multirow{4}{*}{Java}     & C      & \multirow{4}{*}{10.96} & \cellcolor{gray!30} \textbf{37.39} & \multirow{4}{*}{0} & \cellcolor{gray!30} \textbf{5.96} \\
                            &                           & C++    &  & \cellcolor{gray!30} \textbf{20.92} &  & 0 \\
                            &                           & Go     &  & 2.17  &  & 0 \\
                            &                           & Python &  & 3.79  &  & 0 \\ \cline{2-7}
                            & \multirow{4}{*}{Python}   & C      & \multirow{4}{*}{16.23} & \cellcolor{gray!30} \textbf{44.95} & \multirow{4}{*}{0} & \cellcolor{gray!30} \textbf{10.82} \\
                            &                           & C++    &  & 8.89  &  & 0 \\
                            &                           & Go     &  & 2.75  &  & 0 \\
                            &                           & Java   &  & 4.76  &  & 0 \\
\cline{1-7}
EvalPlus & Python & Java & 1.86 & 1.84 & 0 & 0 \\
\hline
TransCoder & Python & Java & 3.53 & 3.52 & 0 & 0 \\
\hline
\end{tabular}
\label{tab:translation_issues}
\end{table}

\subsection*{ RQ3. \RQThreeNew}
\label{rq3}

\pa{Motivation.} \tool{} is composed of several specialized agents and sub-components that operate in sequence: source-to-target translation, \NLSpecS{} generation, \NLSpecS{} validation through pseudocode alignment, test generation, fault localization, and program repair. While RQ1 establishes that the full pipeline outperforms strong baselines across all four benchmarks, it does not reveal where the gains originate. The broad ablation alone does not expose the marginal contribution of every component of the multi-agent framework. This research question therefore proceeds in two complementary parts. We first conduct a broad ablation across multiple datasets, isolating the four functional blocks of the Refinement Agent. We then conduct a deeper, component-level ablation on EvalPlus, in which key sub-components and translation-input configurations, including the Test Agent, the bug-localization-and-repair subsystem, and the prompt-time use of source code versus \NLSpecS{}, are systematically isolated or contrasted under a single, controlled experimental setting.

{\pa{Method.} \textbf{Broad ablation of the Refinement Agent across datasets.} The Refinement Agent comprises four functional blocks: Initial Translation and Fixation (ITF), \NLSpecS{} Augmentation, \NLSpecS{} Validation, and Bug Scope Estimation (BSE). We progressively enable these blocks to obtain four pipeline stages. In Stage~1, the Translation Agent produces a translation that may be refined by the basic Refinement Agent, but \NLSpecS{} generation, \NLSpecS{} validation, and Bug Scope Estimation remain disabled. In Stage~2, \NLSpecS{} is generated and used to guide translation, while Spectrum-Based Fault Localization (SBFL) and the LLM-based repair module are also enabled. In Stage~3, the \NLSpecS{} Validator (Sec.~\ref{nl-validation}) is added. Finally, in Stage~4, Bug Scope Estimation (Sec.~\ref{bug-scope-estimation}) is enabled, yielding the full Refinement Agent. Each stage is evaluated on the test suites included with the benchmarks. Because TransCoder lacks ground-truth evaluation tests, this analysis is conducted on Avatar, CodeNet, and EvalPlus.

\noindent \textbf{Detailed component ablation on EvalPlus.} The broad ablation captures the cumulative effect of the Refinement Agent, but does not isolate the contribution of every individual sub-component, in particular, the Test Agent, the bug-localization-and-repair subsystem, and the role of source code versus \NLSpecS{} at the translation prompt. To address this, we design seven controlled configurations that isolate or contrast key components and translation-input choices within the pipeline. The configurations differ only in the explicitly described component or input selection choices, while the remaining experimental settings are kept identical (GPT 4o, temperature = 0.8). The configurations are defined as follows:

\begin{enumerate}

    \item \textbf{S1. Translate + Compile Fix only} The Translation Agent produces a translation from the source code, the Debugger repairs compilation errors but no test cases are generated or executed. This serves as a lower-bound baseline isolating raw translation and compile-time correctness.

    \item \textbf{S2. + Test-driven Repair} The Test Agent (Sec.~\ref{test-agent}) is enabled. Generated tests drive an iterative repair loop, but \NLSpecS{}, validation, and bug-scope estimation remain disabled. This measures the contribution of test-driven feedback alone.

    \item \textbf{S3. + \NLSpecS{} (no source in prompt)} The Translation Agent receives \NLSpecS{} as input but \emph{not} the source code. \NLSpecS{} Validation is disabled, while bug localization and repair remain active. This configuration isolates the value of natural-language specifications when they replace the source code at the translation step.

    \item \textbf{S4. + \NLSpecS{} Validator} On top of S3, the alignment validation step (Sec.~\ref{nl-validation}) is reintroduced. The source code is still absent from the translation prompt. This measures the contribution of alignment validation in catching specification misalignments before execution.

    \item \textbf{S5. + Source in Prompt} The source code is restored at the translation step, jointly with \NLSpecS{}, but the \NLSpecS{} Validator remains disabled. Bug localization and repair stay active. This isolates whether including the source code at prompt time (in addition to \NLSpecS{}) improves translation.

    \item \textbf{S6. Full minus BugScope} All components are enabled except Bug Scope Estimation. SBFL alone localizes faults, and the Repair Agent operates without the bug-scope classifier. This measures the marginal contribution of Bug Scope Estimation.

    \item \textbf{S7. Full \tool{}} The complete pipeline with all agents, validator, SBFL, and Bug Scope Estimation enabled. This is the reference configuration.

\end{enumerate}

We restrict the detailed ablation to the EvalPlus benchmark (164 instances, Python$\rightarrow$Java). EvalPlus is the most rigorous of our four benchmarks: it ships with extensive functional test suites that catch subtle correctness violations missed by simpler I/O-based test cases, providing the strongest ground-truth evaluation signal for fine-grained component analysis. In addition, running seven configurations across all four benchmarks and every language pair would require roughly an order of magnitude more LLM calls than the broad ablation above. Concentrating the detailed analysis on EvalPlus keeps the experimental cost manageable while preserving statistical meaningfulness, as EvalPlus instances are the most challenging in our suite. To strengthen this analysis, we report two complementary metrics: \CAGen{}, computed on the test cases generated by the Test Agent, and \CAEval{}, computed on the ground-truth evaluation tests shipped with EvalPlus. \CAEval{} provides a fair cross-configuration comparison because every configuration is evaluated against the same fixed test suite. \CAGen{} is reported alongside as the metric on which the internal repair loop operates.
}
\begin{table*}[]
\small
\centering
\renewcommand{\arraystretch}{1.2}
\caption{Translation accuracy for different parts of RA with GPT-4o (\textbf{RA = Refinement Agent, ITF = Initial Translation and Fixation, NL Spec = \NLSpecS augmentation, NL Val. = \NLSpecS validation, BSE = Bug Scope Estimation})}
% \begin{tabular}{|c|c|c|c|c|}
\begin{tabular}{|p{1.2cm}|p{1cm}|p{1.5cm}|p{1.2cm}|p{2cm}|p{2cm}|p{2cm}|}
\hline
\multirow{2}{*}{\textbf{Dataset}} & \multirow{2}{*}{\textbf{Source}} & \multirow{2}{*}{\textbf{Targets}} & \multicolumn{4}{c|}{\textbf{Computational Accuracy}}\\
\cline{4-7}
& & & \textbf{ITF} & \textbf{ITF+NL Spec} & \textbf{ITF+NL Spec+NL Val.} & \textbf{ITF+NL Spec+NL Val.+BSE}\\
\hline
\multirow{8}{*}{Avatar} & \multirow{4}{*}{Java}     & C & 92.1 & 94.16 & 95.41 & 95.83\\
                        &                           & C++ & 93.75 & 95 & 95.83 & 96.25\\
                        &                           & Go & 61.25 & 85.41 & 86.67 & 87.5\\
                        &                           & Python & 88.33 & 92.08 & 92.5 & 92.5\\ \cline{2-7}
                        & \multirow{4}{*}{Python}   & C & 69.87 & 85.41 & 86.25 & 86.67\\
                        &                           & C++ & 79.58 & 87.5 & 87.5 & 87.92\\
                        &                           & Go & 67.08 & 82.92 & 87.17 & 84.17 \\
                        &                           & Java & 84.1 & 87.5 & 87.92 & 88.33\\
\hline

\multirow{20}{*}{CodeNet}   & \multirow{4}{*}{C}        & C++ & 97.5 & 98.5 & 99 & 99\\
                            &                           & Go & 73 & 85 & 86 & 87.5\\
                            &                           & Java & 95.5 & 98 & 98 & 98\\
                            &                           & Python & 88.5 & 92 & 93 & 93\\ \cline{2-7}
                            & \multirow{4}{*}{C++}      & C & 94 & 96 & 96.5 & 98\\
                            &                           & Go & 88 & 92.5 & 93.5 & 94\\
                            &                           & Java & 95.5 & 96.5 & 98 & 98\\
                            &                           & Python & 90.5 & 93 & 94 & 94\\ \cline{2-7}
                            & \multirow{4}{*}{Go}       & C & 90 & 95.5 & 97 & 97.5\\
                            &                           & C++ & 96.5 & 98 & 98.5 & 98.5\\
                            &                           & Java & 94 & 97 & 98.5 & 99\\
                            &                           & Python & 94 & 95.5 & 96 & 96.5\\ \cline{2-7}
                            & \multirow{4}{*}{Java}     & C & 92 & 95 & 97 & 97.5\\
                            &                           & C++ & 94.5 & 96 & 97.5 & 98\\
                            &                           & Go & 63 & 86.5 & 87.5 & 87.5\\
                            &                           & Python & 88.5 & 92 & 92.5 & 94\\ \cline{2-7}
                            & \multirow{4}{*}{Python}   & C & 96.5 & 97 & 98 & 98\\
                            &                           & C++ & 97.5 & 98.5 & 99 & 99.5\\
                            &                           & Go & 81.5 & 93.5 & 93.5 & 94\\
                            &                           & Java & 98 & 99 & 99 & 99\\
\hline
EvalPlus & Python & Java & 86.59 & 87.8 & 89.63 & 91.46\\
\hline
\multicolumn{3}{|c|}{Average} & 87.28 & 92.85 (+5.57\,pp) & 93.72 (+0.87\,pp) & 94.18 (+0.46\,pp)\\
\hline
\end{tabular}
\label{tab:ablation_study}
\end{table*}

\begin{table}
\centering

\caption{Detailed ablation on EvalPlus (Python$\rightarrow$Java, GPT-4o, $N=164$), showing cumulative accuracy: the fraction of instances solved by that stage or any earlier stage in the sequence S1–S7. \CAGen{}: accuracy on Test-Agent-generated test cases, \CAEval{}: accuracy on the ground-truth EvalPlus test suite.}
\label{tab:rq3_ablation}
\begin{tabular}{clcc}
\toprule
\textbf{Stage} & \textbf{Component Added}
  & \textbf{\CAGen{} (\%)}
  & \textbf{\CAEval{} (\%)} \\
\midrule
S1 & Translate + Compile Fix only             & --    & 79.88 \\
S2 & + Test-driven Repair                     & 84.76 & 85.37 \\
S3 & + \NLSpecS{} (no source in prompt)       & 89.63 & 88.41 \\
S4 & + \NLSpecS{} Validator                   & 91.46 & 89.63 \\
S5 & + Source in Prompt                       & 92.68 & 90.24 \\
S6 & + Full pipeline minus BugScope           & 93.29 & 90.24 \\
\midrule
S7 & + BugScope Estimation (Full \tool{})     & \textbf{93.29} & \textbf{91.46} \\
\bottomrule
\end{tabular}
\end{table}

\begin{figure}
    \centering
    \includegraphics[width=\textwidth]{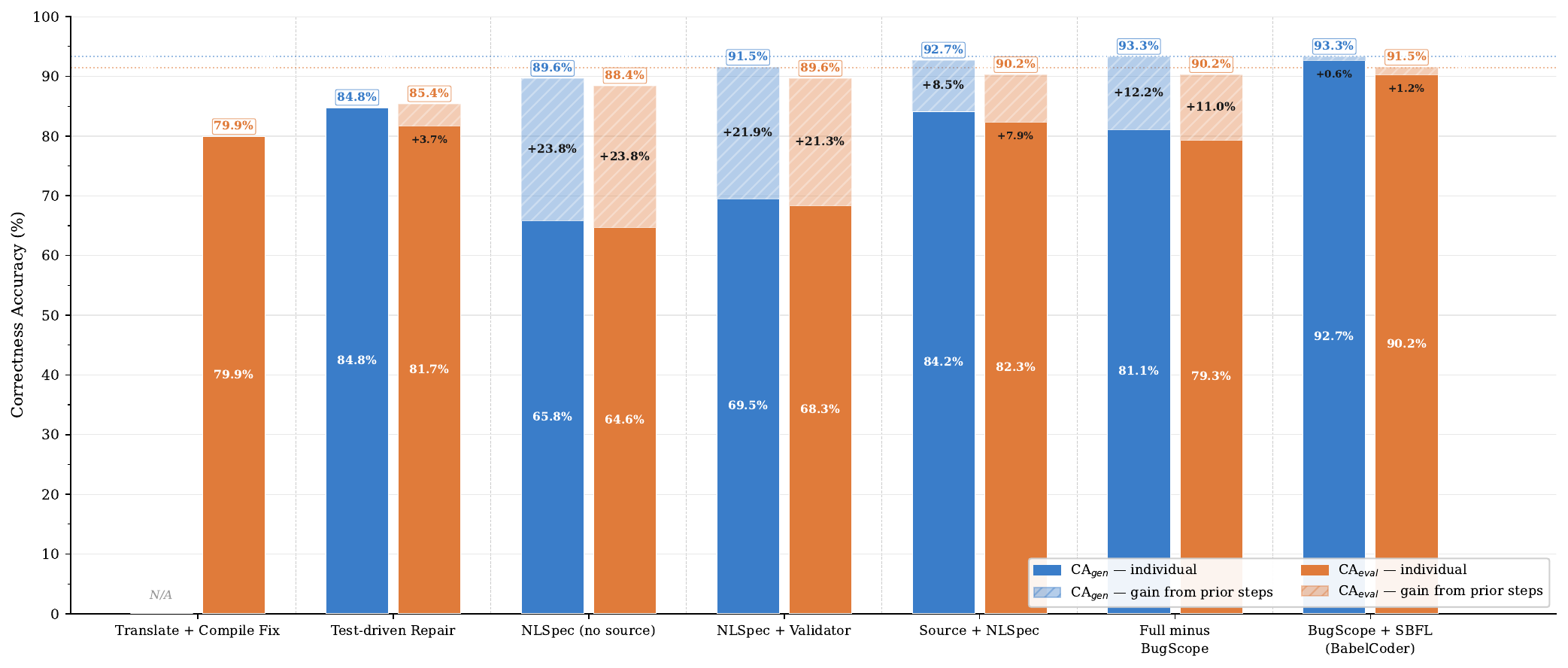}
    \caption{Detailed ablation study on EvalPlus (Python$\rightarrow$Java, GPT-4o, $N=164$). Each pair of bars corresponds to one of the seven configurations (S1-S7). The blue bar reports \CAGen{} and the orange bar reports \CAEval{}. The solid lower segment of each bar is that configuration's accuracy on its own. The hatched upper segment adds instances solved only by an earlier configuration. The full bar height is the cumulative accuracy reported in Table~\ref{tab:rq3_ablation}, and the dotted line marks the cumulative accuracy of the full pipeline (S7).}
    \label{fig:ablation}
\end{figure}

{\pa{Result.} \textbf{Broad ablation of the Refinement Agent across datasets.} Table~\ref{tab:ablation_study} reports the broad ablation across Avatar, CodeNet, and EvalPlus, covering 29 source-target language pairs. At Stage~1, where the Refinement Agent operates without \NLSpecS{}, validation, or Bug Scope Estimation, the average CA is 87.28\%. Adding \NLSpecS{} Augmentation in Stage~2 raises the average CA to 92.85\%, a gain of 5.57 absolute percentage points (pp), with the largest individual improvements observed on Avatar Java$\rightarrow$Go (+24.16\,pp), CodeNet Java$\rightarrow$Go (+23.5\,pp), and Avatar Python$\rightarrow$C (+15.54\,pp), all cases in which the source and target languages differ substantially in syntax and idiom. \NLSpecS{} Validation (Stage~3) brings a further 0.87\,pp improvement to 93.72\%, and Bug Scope Estimation (Stage~4) yields the final 94.18\%. Every component of the Refinement Agent therefore contributes a measurable improvement, and the gains compound consistently across datasets and language pairs.

\noindent \textbf{Detailed component ablation on EvalPlus.} Table~\ref{tab:rq3_ablation} and Figure~\ref{fig:ablation} present the detailed component ablation on EvalPlus. We make three observations.

\begin{enumerate}
    \item \textit{Test-driven repair is the largest single contributor.} Moving from S1 (Translate + Compile Fix only) to S2 (+ Test-driven Repair) raises the cumulative CAeval from 79.88\% to 85.37\%, a gain of 5.49 pp. This confirms that the Test Agent's iterative repair loop is the most influential single component of the pipeline, recovering a substantial number of translations that fail purely at runtime rather than at compile time.
    \item \textit{Source code at the prompt is essential. \NLSpecS{} alone is insufficient.} The individual results in Figure~\ref{fig:ablation} show that the \NLSpecS{}-only configurations (S3, S4), in which the Translation Agent does not see the source code, achieve only 64.63\% and 68.29\% \CAEval{} respectively, well below the 81.71\% obtained by translation from source code alone (S2). The source code therefore remains a necessary signal for the translation step. When source code is reintroduced alongside \NLSpecS{} in S5, the individual \CAEval{} rises to 82.32\%, surpassing source-only translation and demonstrating that source and \NLSpecS{} are complementary rather than substitutable. The sequential view in Table~\ref{tab:rq3_ablation} reinforces this: the cumulative \CAEval{} progresses smoothly from 85.37\% at S2 to 90.24\% at S5, with each stage adding non-trivial new coverage.
    \item \textit{\NLSpecS{} Validation and Bug Scope Estimation provide measurable but smaller gains.} Adding \NLSpecS{} Validation (S4) over S3 yields 1.22\,pp of cumulative \CAEval{} improvement, and adding Bug Scope Estimation as the final component (S7 over S6) recovers a further 1.22\,pp, bringing the cumulative \CAEval{} from 90.24\% to 91.46\%. While each gain is individually modest, both components consistently recover instances that earlier stages cannot, and together they account for translations that the test-driven repair loop alone is unable to fix. Through S7, the cumulative CAgen reaches 93.29\% and cumulative CAeval reaches 91.46\%, the number marked by the dotted line in Figure 9 and the same accuracy reported for the full pipeline elsewhere in the paper (Table 2, Table 5), confirming that every component of BabelCoder makes a measurable and complementary contribution to overall translation correctness. The S7 configuration's own accuracy, run on its own without crediting instances recovered only by weaker stages, is the solid segment of the final bar in Figure 9.
\end{enumerate}

\smallskip
\noindent\textbf{Takeaway.} Across both ablations, the test-driven repair loop and \NLSpecS{}-guided translation are the two most impactful components, jointly delivering most of the improvement over a raw translation baseline. \NLSpecS{} Validation and Bug Scope Estimation contribute smaller but consistent additional gains, and the source code remains an essential input even when \NLSpecS{} is available. The complementary nature of these components justifies the multi-agent design of \tool{}.}

\section{Related Work}
\noindent Recent advancements in large language models (LLMs) have significantly impacted software engineering tasks, particularly code translation and code generation. We review related work in two areas: (1) \textbf{Code Translation}, which converts code between languages, and (2) \textbf{Code Generation}, which creates code from high-level specifications.

\subsection{Code Translation}
\noindent In recent years, several approaches have been proposed for function-level and file-level code translation.

\pa{Function-Level Code Translation}.
Previous efforts in code translation with LLMs have largely focused on function-level translation. To improve translation capabilities, several approaches enhance LLMs through model design and contextual augmentation. For example, SteloCoder \citep{pan2023stelocoderdecoderonlyllmmultilanguage} adapts the StarCoder architecture into a decoder-only LLM for multi-language code translation. SpecTra \citep{nitin2024spectraenhancingcodetranslation} incorporates multi-modal specifications, i.e., natural language descriptions, control flow graphs, and data flow graphs, to guide translation. AGL-Code \citep{yu2023pseudocode} adaptively leverages global and local contextual information to translate pseudocode into executable code, while Bhattarai et al.~\citep{bhattarai2024enhancing} improves few-shot learning through retrieval-augmented generation, injecting relevant translation examples to guide the LLM. Saha et al.~\citep{saha2024specification} use NL-Specification as an intermediate representation while translating code from source to target language.

Several works also focus on translation correction and functional validation. Rectifier \citep{yin2024rectifiercodetranslationcorrector} learns from faulty-corrected code pairs to repair common translation errors. TransAgent \citep{yuan2024transagentllmbasedmultiagentcode} employs a multi-agent system responsible for test generation, syntax error repair, and semantic alignment, using source-target execution behavior for automated translation correction. Similarly, UniTrans \citep{yang2024exploring} leverages test case generation and iterative repair to improve translation accuracy, while ExeCoder \citep{he2025execoder} incorporates semantics and dependency information to improve executability. CoTR \citep{yang2025assessing} enhances robustness via syntactic transformations and adversarial training.

Finally, for empirical analysis, LIT \citep{pan2024lost} systematically studies bugs introduced by LLMs during code translation, categorizing common failure modes and proposing targeted evaluation metrics. Complementing model-centric analyses, Rabbi et al.~\citep{rabbi2026beyond} show that code-translation evaluations can also report false failures caused by compilation and runtime configuration rather than incorrect translation logic, highlighting the importance of reliable evaluation infrastructure.

\pa{Repository-Level Code Translation}. In addition, several works focus specifically on repository-level code translation. K3Trans \citep{ou2025enhancing} is a knowledge-augmented LLM approach that leverages target code samples, dependency usage, and historical translations to enhance translation quality and correctness in complex repository contexts. AlphaTrans \citep{ibrahimzada2024alphatrans} adopts a neuro-symbolic method that uses static analysis to decompose programs and translates code fragments in reverse call order, supporting both code and test translation with multi-level validation. To aid evaluation efforts of such systems, RepoTransBench \citep{wang2024repotransbench} introduces a benchmark of 100 real-world repositories with automated test suites, highlighting the challenges current LLMs face in repository-level code translation and exposing common failure modes.

While prior works have explored LLM-based code translation using prompt engineering, retrieval augmentation, or fine-tuned models, they often treat translation and error correction as separate tasks or rely on single-pass generation. Multi-agent frameworks demonstrate the potential of collaboration but are typically limited to either translation or repair. In contrast, our work introduces an end-to-end multi-agent framework that seamlessly integrates translation, bug localization, and targeted repair within a unified pipeline. The system leverages both source code and intermediate natural language specifications (\NLSpec) to guide translation, with validation mechanisms for \NLSpecS and novel techniques for error handling, including LLM-based scope estimation, spectrum-based fault localization, and LLM-guided repair using localized bug information.

\subsection{Code Generation}
Recent work on code generation emphasizes repository-level generation, aiming to produce accurate and functional code across entire software projects while accounting for complex dependencies and context. This can be achieved using either single-agent or multi-agent approaches.
In single-agent approaches, a single LLM is responsible for handling the entire generation task. For example, CatCoder \citep{pan2024enhancing} incorporates structural code context to LLM prompts, showing that enriching prompts with repository-specific structural data significantly improves single-agent-based code generation. In contrast, TOOLGEN \citep{wang2024teaching} integrates auto-completion tools into LLM-based code generation to better mimic developer-tool interaction.
On the other hand, multi-agent approaches divide responsibilities among multiple agents. Self-collaboration \citep{dong2024self} uses role-based agents to simulate a structured software team dynamics to improve performance, or self-organized agents \citep{ishibashi2024self}, each autonomous agent would handle sub-components independently, then collaborate to scale on large projects. CodeAgent \citep{zhang2024codeagent} combines multiple specialized agents with external coding tools, enabling realistic, tool-based code generation.
Complementing these methods, HALLUCODE \citep{liu2024exploring} introduces a benchmark to evaluate LLMs' ability to detect and mitigate hallucinations, revealing current models’ limitations in handling such issues.

These methods are closely related to code translation, particularly when pseudocode or specifications are used as in \tool. Aside from the obvious difference in the final objective, the key distinction is that \toolS leverages both source code and accompanying specifications to guide the translation process.

Recent work also examines quality dimensions beyond correctness, including robustness across languages and requirements \citep{rabbi2025multi, lin2025robunfr} and, through HEJ-Robust, robustness of LLM-based program repair under semantics-preserving transformations \citep{rabbi2026hejrobust}. Bias Unveiled and a subsequent SocialBias-Bench study analyze social bias and mitigation in generated code \citep{ling2025bias, rabbi2026socialbias}, while fine-tuning and synthesized secure instruction data have been explored for secure code generation \citep{li2024exploratory, li2025prompt}. Complementing these quality-oriented studies, Mantra \citep{xu2025mantra} explores automated method-level refactoring using contextual retrieval-augmented generation and multi-agent LLM collaboration, extending LLM applications to software maintenance tasks.

\section{Threats to Validity}
\label{threats}

While \tool{} demonstrates strong performance across multiple datasets and language pairs, several factors may influence the generalizability and consistency of our results. We outline key threats below.

\pa{External Validity.}
A primary threat is the potential overlap between benchmark datasets and the training data of the underlying LLM. Such overlap may inflate performance if the model has been previously exposed to similar code or problem patterns. Furthermore, the proprietary and opaque nature of modern LLMs limits transparency regarding their training data and internal mechanisms, making it difficult to assess hidden biases, memorization effects, or systematic failure modes. These factors may restrict the extent to which our findings generalize to unseen or real-world codebases.

\pa{Internal Validity.}
Our results may be affected by the nondeterministic behavior of LLMs. Even with fixed prompts and identical inputs, repeated invocations can produce slightly different translations or repairs, introducing minor variations in outcomes. In addition, our evaluation depends on automatically generated or converted test suites. In particular, the static converter used to transform PyTest cases into JUnit tests may fail on rare or complex input types, potentially affecting correctness assessments. Although such cases were infrequent, they may influence individual evaluations. Improving the robustness of this converter is part of our future work.

{Finally, we note that the oracles used inside the pipeline are produced by running the source program on the generated inputs, and the benchmark's ground-truth evaluation tests are held out and never seen by any agent during translation or repair. The repair signal and the evaluation signal are therefore disjoint, which rules out any leakage from the evaluation oracle into the reported results.}

{\pa{Construct Validity.} Our quality and security analysis in RQ2 relies on SonarQube, a static analyzer that can report false positives, and whose false-positive rate varies across languages. We therefore interpret its output as a relative indicator of the change between the source and the translated code under an identical configuration, rather than as an absolute or definitive measure of security. Corroborating these findings with dynamic analysis or manual auditing is left for future work.}
\section{Conclusion and Future Work}
This paper presents \tool, a multi-agent autonomous code translation tool that treats translation as a distributed problem, solved collaboratively by multiple agents with the assistance of large language models (LLMs). \tool{} comprises three specialized agents, each responsible for a distinct sub-task in the translation pipeline. It introduces several novel techniques, including validating \NLSpecS using an LLM-based aligner, localizing bugs with an LLM-driven scope finder, generating precise error messages, and repairing bugs using location-aware prompts. 

\tool{} has been evaluated on four public benchmark datasets across five programming languages and compared against four strong baselines. Experimental results show that \tool{} consistently outperforms existing approaches across nearly all language pairs, achieving higher accuracy and robustness, handling diverse programming languages, varying code complexities, and real-world input variations without significant performance degradation.

As future work, we aim to enhance \tool’s capabilities by incorporating control flow information from the source code, enabling more accurate bug localization and further improving the overall quality of translations.

\clearpage
% \bibliographystyle{abbrv}
% \bibliography{references}

\end{document}